\documentstyle[aps,epsf,preprint]{revtex}
\begin{document}
\title{Onset of Propagation of Planar Cracks in Heterogeneous Media}
\author{Sharad Ramanathan
\thanks{
Present Address: Institute for Theoretical Physics, University of
California, Santa Barbara, CA 93106-4030.} 
and Daniel S. Fisher}
\address{Lyman Laboratory of Physics, Harvard University, Cambridge,\\
Massachusetts 02138}
\date{\today}
\maketitle

\begin{abstract}
The dynamics of planar crack fronts in hetergeneous media near the
critical load for onset of crack motion are investigated both
analytically and by numerical simulations. Elasticity of the solid
leads to long range stress transfer along the crack front which is
non-monotonic in time due to the elastic waves in the medium. In the
quasistatic limit with instantaneous stress transfer, the crack front
exhibits dynamic critical phenomenon, with a second order like
transition from a pinned to a moving phase as the applied load is
increased through a critical value. At criticality, the crack-front is
self-affine, with a roughness exponent $\zeta =0.34\pm 0.02$. The
dynamic exponent $z$ is found to be equal to $ 0.74\pm 0.03$ and the
correlation length exponent $\nu =1.52\pm 0.02$. These results are in
good agreement with those obtained from an epsilon expansion.
Sound-travel time delays in the stress transfer do not change the
static exponents but the dynamic exponent $z$ becomes exactly
one. Real elastic waves, however, lead to overshoots in the stresses
above their eventual static value when one part of the crack front
moves forward. Simplified models of these stress overshoots are used
to show that overshoots are relevant at the depinning transition
leading to a decrease in the critical load and an apparent jump in the
velocity of the crack front directly to a non-zero value. In finite
systems, the velocity also shows hysteretic behaviour as a function of
the loading. These results suggest a first order like
transition. Possible implications for real tensile cracks are
discussed.
\end{abstract}


\section{Introduction}

The dynamics of cracks in heterogeneous media is a very rich field
involving much physics that is yet to be understood. Even in
situations in which the {\it path} of a crack is predetermined -- for
example by a pre-weakend fault-- its dynamics can still be
complicated. The simplest situation is a crack confined to a
plane. For small loads across such a planar crack, the crack front
will be at rest. As the load is gradually increased, the crack front
may undergo some transient motion but then again be arrested. If the
load is increased above a critical load, however, the crack front will
begin to propagate through the sample. The behaviour near to the {\it
onset} of propagation of planar cracks -- in particular tensile cracks
-- is the subject of this paper.

In recent years there has been considerable theoretical progress
towards understanding the dynamics of elastic manifolds moving through
random media, such as charge density waves\cite{on}, fluid--surface
contact lines\cite{de} and interfaces between two phases. All of these
exhibit a type of non-equilibrium critical phenomenon near to the
onset of motion. However there are various features which make the
system of a planar crackfront moving through a heterogeneous medium
different from these other systems.

For cracks (as well as for contact lines) the bulk degrees of freedom
lead to effective long range interactions between the points on the
front\cite{degennes}\cite{ricelongrange}. Thus, when a point on the
crack front moves ahead, the stress at all other points on the front
increases due to the elastic interactions tending to pull them
forward. In addition, elastic waves are emitted as the crack front
moves non-uniformly. When one point moves ahead, these waves result 
in
stresses elsewhere on the front which, for a while, are greater than
those due to just the static elastic deformations which will obtain
long after the waves have passed. Both these {\it stress overshoots}
and the long range interactions have earlier been shown to play a
crucial role in the dynamics of the crack front when it is moving with
a non-zero mean velocity.\cite{perrin}\cite{dynamicsdsf}

In the {\it absence} of these stress overshoots --- as obtains if the
stress transfer is quasistatic --- many aspects of the dynamics of a
planar crack front near the onset of motion can be understood by
analogy with interfaces, in particular via a functional
renormalization group analysis, which for cracks, as for contact
lines, entails an expansion about two dimensions\cite{de}. The
phenomenology is built on the existence of two ``phases'' which are
separated by a {\it unique} critical load. When the applied load,
$G^{\infty }$, is small, there is no steady-state motion and the crack
front is pinned by the random toughness in one of many locally stable
configurations--- we will ignore here and henceforth the effects of
thermal creep. As the load is increased adiabatically, there are a
series of local instabilities of the crack front which lead to
``avalanches'' that can become large as $G^{\infty }$ is increased
further. Eventually at the { critical} load, $G_{c}^{qs}$, the crack
front de-pins and begins to move --- albeit very jerkily --- with a
non-zero, mean steady state velocity, $v$. In an infinite system, this
transition from the stationary to the moving phase exhibits
non-equilibrium dynamic critical phenomena somewhat analogous to 
those
near conventional second-order transitions. One macroscopic
manifestation of this is the behaviour of the mean velocity of the
crack front at a load just above the critical load:
\begin{equation}
v\sim (G^{\infty }-G_{c}^{qs})^{\beta }.  \label{velocity}
\end{equation}

A natural question that arises is the role of the stress overshoots
left out of the quasistatic analysis. In particular, what are their
effects on the crack dynamics and how do these affect the depinning
transition? The temporal shape of the stress overshoots seen by a
point on the crack front depends on various microscopic details, such
as the microscopic response time of the crack front, acoustic damping
processes etc. How the dynamics of the crack front depends on the
nature of the stress overshoots and if there is any limit in which the
dynamics of the medium can be neglected are not understood; these are
questions that must be addressed. In particular, in the presence of
stress overshoots, is there a regime in which a second order like
transition from the pinned to the moving state persists or does the
crack front always jump directly to a finite velocity? If the stress
overshoots are ``relevant'' at the depinning transition do they make
the velocity versus loading curves hysteretic? In either case, is the
`moving phase' just above the threshold a non-trivial statistically
stationary state or is it characterized by noisy linear dynamics?
Thus, there are a large number of unanswered questions even in the
seemingly simple problem of the dynamics, near threshold, of a crack
front restricted to move in a plane.

In this paper we study the dynamics of a crack front restricted to
move in a plane, through a three dimensional solid with
heterogeneities only in the local fracture toughness. The effects of
both the long range interactions and the stress pulses are considered,
and some of the questions raised above addressed. In the absence of
the stress overshoots, we obtain, numerically, some of the exponents
which characterize the transition from the stationary to the moving
phase, check the scaling laws that have been predicted and compare the
exponents with the analytical results obtained earlier by the $
2-\varepsilon $ expansion\cite{de}. We then extend the analysis to
include the effects of sound travel time delays in the stress
transfer. Finally, we treat the effects of the stress overshoots on
the depinning transition. Both the dynamic stresses obtained from a
scalar approximation to elasticity and sharp pulse-like overshoots are
studied.

\subsection{Outline} 

Before introducing the basic model and summarizing our main results,
we give an outline of the paper. In section {\ref{section2}} the
details of the models and the numerical methods employed are
described. Section {\ref{quasistaticr}} contains the results of the
quasi-static model, where the stress transfer is instantaneous, while
Section \ref{timedelayr} contains those in the case where there are
acoustic time delays in the stress transfer. In Section \ref
{nonmonor}, the effects of various kinds of stress overshoots are
explored.  Finally the results and their possible implications are
discussed in section \ref{discussion3}. The ``no-passing
rule''\cite{nopassing} for these models, which plays an essential role
in the analytical results, is discussed in Appendix \ref{nopass}. Appendix 
\ref{kernelsnum} has
the detailed forms of the kernels used in the numerical simulations.

\subsection{Summary of Results}

The equation of motion for the crack front can be obtained by
requiring energy conservation at all points on the crack front. This
implies that the elastic energy flux into the crack, which is a
non-local functional in both space and time of the shape of the crack
front as well as of the local velocity of the crack front, must be
equal to the surface energy required to create the new crack surface,
i.e., the {\it local fracture toughness}. The general {\it linearized}
equation of motion for a crack front moving along the positive $x$
direction has the form,
\begin{equation}
\partial _{t}f(z,t)=\int\limits_{t^{\prime }<t}dt^{\prime }{\cal P}\int
dz^{\prime}J(z-z^{\prime },t-t^{\prime })\partial _{t^{^{\prime
}}}f(z^{\prime },t^{\prime })-\gamma [f(z,t),z]+{\cal E}  \label{eqm}
\end{equation}
where $z$ is the co-ordinate along the crack front, ${\cal P}$ denotes
the principal part of the integral, $f(z,t)$ is the deviation of the
crack front from a straight one, $\gamma $ is a random variable
associated with the random position dependent fracture toughness in
the solid and ${\cal E}$ represents the driving ``force'' due to the
applied load, $G^{\infty}$. The kernel $J$ is non-local both in space
and time. This non-locality arises from the long range elastic
interactions and the sound waves which are emitted as the crack
moves. Note that because the basic processes near threshold consist of
sections of the front moving ahead and stopping --- i.e., roughly step
functions in time --- we have chosen to write the stress transfer in
terms of $\partial _{t^{^{\prime }}}f(z^{\prime },t^{\prime })$, so
that these jumps are approximately delta functions in $t^{\prime }.$

We will classify the models based on whether or not the kernel, $J$,
is {\it monotonic} in time at every spatial co-ordinate,
$z$. Monotonicity of the stress transfer plays a crucial role. It
means that as a segment of the crack moves forward, the stress at all
other points increases monotonically in time.  This {\it convexity}
property yields stringent constraints on the behaviour as shown in
Appendix \ref{nopass}.
 It implies that a configuration of the crack which is behind
another configuration at one time will remain behind the other
configuration at all later times. This immediately leads to the
conclusion that there is a unique critical load, $G_{c}^{qs}$ for
monotonic models.

\subsubsection{Quastatic Approximation}

We first consider the quasistatic approximation in which sound waves
are neglected and the stress transfer is {\it instantaneous} so that
the kernel is naturally monotonic. In this case, the basic
phenomenology is well known\cite {on,bak}. As the load, $G^{\infty },$
is gradually increased, segments of the crack front will overcome the
local toughness and jump forwards, perhaps causing other segments to
jump and thereby triggering an avalanche which will eventually be
stopped by tougher regions. We find that, similar to driven interfaces,
etc.\cite{on,de}, the avalanches show a power law size distribution up
to a characteristic length, $\xi _{-},$ with larger avalanches being
much rarer. The distribution of avalanche size --- roughly the extent
along the crack front of an avalanche--- has the form
\begin{equation}
{\rm {Prob}}({\rm size\ of\ avalanche }>l)\approx \frac{1}{l^{\kappa }}\hat{
\rho}(l/\xi _{-})  \label{avascaling}
\end{equation}
The cutoff length, $\xi _{-}$, defines the correlation length below
threshold. As the threshold load, $G_{c}^{qs }$, is approached, the
correlation length, $\xi _{-}$ diverges as
\begin{equation}
\xi _{-}\sim (G_{c}^{qs}-G^{\infty })^{-\nu _{-}}.
\end{equation}
At the threshold, there is no characteristic length scale and the
distribution of the avalanche sizes is a pure power law. From Eq.(\ref
{avascaling}) and scaling relations between the exponents, we expect
that the cumulative probability of the size of an avalanche being
greater than $ l, $ as the load is swept slowly from zero to the
critical load scales as
\begin{equation}
\int\limits_{0}^{G_{c,qs}^{\infty }}dG^{\infty }\frac{1}{l^{\kappa }}\hat{
\rho}(l/\xi _{-})\sim \frac{1}{l};
\end{equation}
this is in agreement with the numerics within error bars.

As the load increases above the critical load, the crack front begins
to move with a mean velocity, which the monotonicity implies is
unique. The velocity scales as in Eq.(\ref{velocity}), with the
velocity exponent, determined from our numerical simulations,
\begin{equation}
\beta =0.68\pm 0.06.
\end{equation}
All quoted error bars here and henceforth are one-$\sigma$ errorbars 
from $\chi^{2}$ fits.
Just above $G_{c}^{qs}$ the motion of the front is very jerky with
fluctuations in the velocity correlated up to a distance $\xi _{+}$,
which diverges as one approaches the threshold from above as $\xi
^{+}\sim (G^{\infty }-G_{c}^{qs})^{-\nu _{+}}$.

The exponents $\nu _{+}$ and $\nu _{-}$ will be equal i.e., $\nu
_{+}=\nu _{-}=\nu $, if there is only one divergent length scale in
the problem, as predicted by the renormalization (RG) analysis
\cite{on}. Assuming this two sided scaling, we can obtain the
correlation length exponent, via finite size scaling, from the
dependence of the variance of the critical load on the size of the
system as
\begin{equation}
\nu =1.52\pm 0.02.
\end{equation}

At threshold, the crack front is self-affine with correlations

\begin{equation}
\left\langle \lbrack f(z,t)-f(z+r,t)]^{2}\right\rangle \sim r^{2\zeta }
\end{equation}
where $\left\langle {}\right\rangle $ denotes the average over the
randomness. The roughness exponent $\zeta $ is found numerically to be 
\begin{equation}
\zeta =0.34\pm 0.02,
\end{equation}
in excellent agreement with the 2$-\varepsilon $ expansion prediction 
$\zeta
\approx 1/3.$

The dynamic exponent is found from the duration of avalanches as a
function of size $l$; they typically last for
\begin{equation}
\tau _{l}\sim l^{z}
\end{equation}
with 
\begin{equation}
z=0.74\pm 0.03.
\end{equation}
The exponent identities predicted from the scaling and RG analysis 
\cite{on,de}
\begin{eqnarray}
\beta &=&(z-\zeta )\nu  \label{expid1} \\
\nu &=&\frac{1}{1-\zeta }  \label{expid2}
\end{eqnarray}
are found to be satisfied, so that there are only {\it two independent
exponents}, say $\zeta $ and $z$, characterizing the transition.

\subsubsection{ Effects of Sound Travel Time.}

For a model with a monotonic kernel but with the stress transfer
delayed by the sound travel time, we argue that the static exponents
$\nu $ and $\zeta $ are {\it identical} to the corresponding
quasistatic case. Also, for every manifestation of the randomness, the
critical load for this model, $ G_{c}^{td},$ is exactly equal to that
in the quasistatic approximation, $ G_{c}^{qs}.$ However, the dynamic
exponent for this model is predicted to be $z=1$ exactly. Since the
exponent identities Eq.(\ref{expid1}) and Eq.(\ref{expid2}) also
hold for this model, we obtain $\beta =1,$ which is consistent with
the numerical results.

\subsubsection{Sound Waves and Stress Overshoots.}

The inclusion of the effects of sound waves leads to non-monotonic
kernels.  These result in the stress at points on an advancing crack
front being non-monotonic in time, which substantially changes the
physics. We have considered two types of non-monotonic kernels, the
first arises from a scalar approximation to elasticity theory and the
second is a simpler one characterized by sharp pulses superimposed on
the time-delayed stress transfer. In both cases we find that the
overshoots in the stress are{\it \ relevant} at the depinning
transition, causing large avalanches to run away and changing the
nature of the transition from the pinned to the moving phase.

The model with sharp pulses involves non-monotonic kernels of the form

\begin{equation}
J_{{\rm {sp}}}(z,t;\alpha ,\gamma )=\Theta (t-|z|)/\pi z^{2}+\alpha \delta
(t-|z|)/|z|^{\gamma }  \label{sharppulsekernel}
\end{equation}
For $\alpha =0$, there are no stress pulses and the model reduces to
the sound travel-time delayed model and hence $G_{c}(\alpha =0,\gamma
)$ is identical to the threshold force for the quasistatic model,
$G_{c}^{qs}.$ We find, both from analytic arguments and from the
numerics, that for small positive $\alpha $ and fixed $\gamma \geq
1/2$, the threshold load, $ G_{c}(\alpha ,\gamma )$, decreases with
increasing $\alpha $ as

\begin{equation}
\langle G_{c}^{qs}-G_{c}^{\infty }(\alpha ,\gamma )\rangle \sim \alpha 
^{2}.
\end{equation}
This behaviour is controlled by the relevant eigenvalue for the
overshoot perturbation at the quasistatic depinning fixed point.

In a scalar approximation to elasticity theory, the stress overshoots
have long tails in time. In addition, the rough crack front will affect the
propagation of the stress pulses due to the non-linearities neglected
in Eq.(\ref{eqm}). We argue that the basic features found in the sharp
stress pulse model still obtain, in particular that the stress
overshoots are {\it relevant }and change the nature of the
transition. Numerical results using an appropriate class of kernels
support this conclusion.

For real elastodynamics appropriate to a tensile crack, the stress
transfer kernel, for a fixed $z,$ is found to be initially negative,
when the longitudinal sound waves arrive, and then change sign when
the Rayleigh waves arrive. The stress peaks due to the Raleigh waves
are similar to those in the scalar elastic approximation and we
believe that they will have similar effects in decreasing the critical
load. However, the more complicated nature of the stress transfer
suggests that the depinning transition of tensile cracks may involve
essential additional physics. Some tentative ideas in this direction
are discussed at the end of the paper.

With or without the additional complications of the full elastodynamic
stress transfer, the nature of the transition between a static and a
moving crack front in the presence of stress overshoots is not
resolved by our numerical or analytical results. The simplest
scenario, which appears to be supported by the numerics, is a
``first-order'' transition with hysteresis from a pinned phase to a
state with non-zero velocity. This may well be the correct scenario,
but possible concerns and other possibilities are discussed in section
V.

\section{ Models}

\label{section2} 
In this section we discuss the equation of motion for a
real tensile crack and various approximations to it that we will study.

\subsection{Geometry and Equation of Motion}

\label{eqmotion} 
We denote the plane in which the crack is confined $ y=0,$ with
the crack open in the region $x<F(z,t)$. We assume that $
F(z,t)$ is a single valued function of $z$ so that, the curve
$x=F(z,t)$ describes the location of the crack front. Since the crack
is planar, the fracture surfaces that it leaves behind are, of course,
smooth. The geometry is shown in Fig.[\ref{4crackgeom}].

The vectorial displacement field $\vec{u}$, satisfies the equations of
elastodynamics

\begin{equation}
\rho \partial _{t}^{2}u_{i}=\partial _{j}\sigma _{ij}  \label{elastodyn}
\end{equation}
with $\sigma _{ij}$ the stress tensor. The displacement field,
$\vec{u} (x,y=0^{\pm },z),$ has a discontinuity across the crack
surface while the normal stresses, $\sigma _{iy}(y=0^{\pm })$, must
vanish on the crack surface. For a crack with purely {\it {tensile
loading}}, only $u_{y}$ will be discontinuous and will have a
$\sqrt{F(z,t)-x}$ singularity at the crack front with an amplitude
proportional to the local mode I stress intensity factor $K_{I}(z,t)$
\cite{freund}. As long as the crack remains planar, symmetry under
$y\rightarrow -y$ implies that the loading is purely mode I, so that
we will simply use $K\equiv K_{I}$\cite{quasistatic}. We consider the
system under a static load applied far away so that for a straight
crack at rest (i.e., $F(z,t)=const$), $K=K^{\infty }=const.$

As the crack front advances, $F\rightarrow F+\delta F$, an energy per
unit area of the new crack surfaces exposed, $\Gamma [x=F(z,t),z],$
must be provided to the crack front in order to fracture the solid; in
an ideal quasi-equilibrium situation this is just twice the
solid--vacuum interfacial energy density, more generally it is the
local fracture toughness that includes the effects of small scale
physics for which linear continuum elasticity is not valid. The
fracture energy will be provided to the crack-front by a flux of
stored elastic energy per unit area of the new crack surface, ${\cal
G}(z,t,\{F\})$, which in general depends on the past history of the
whole crack front as well as its instantaneous local velocity
$\frac{\partial F}{\partial t}$ . The equation of motion of the crack
front is obtained by requiring that the elastic energy released be
equal to the surface energy required for fracture, i.e.,

\begin{equation}
{\cal G}[z,t,\{F(t^{^{\prime }}\leq t)\}]=\Gamma [x=F(z,t),z]
\label{generaleqn}
\end{equation}
for all $z$ and $t$. The available energy , ${\cal G}$, has the general form

\begin{equation}
{\cal G}={A}[v_{\bot }(z,t)]G[z,t\{F(t^{^{\prime }}<t)\}]  \label{gstructure}
\end{equation}
where $v_{\bot }$ is the local velocity normal to the crack front and
$G$, which is independent of $\frac{\partial F}{\partial t}(z,t)$, is
the elastic energy that would be released at $(z,t)$ if the crack had
advanced {\it adiabatically} at that point, i.e., with $\frac{\partial
F}{\partial t} (z,t)=0 $ \cite{thesis}.

For a straight stationary crack, $F=const$, 
\begin{equation}
G=G^{\infty }=\frac{1-\nu^{2}}{E}(K^{\infty })^{2},
\end{equation}
with $E$ the Young's Modulus and $\nu $ the Poissons
ratio\cite{lawn}. When the crack advances at a non-zero velocity, not
all of the released elastic energy is available for fracture; some
fraction of it goes into the kinetic energy of the moving material
very close to the front. The fraction of $G$ available for fracture
$A[v_{\bot }(z,t)]$ depends {\it only} on the local normal velocity;
it decreases from unity for small ${v}_{\bot }$ and goes to zero for
${v}_{\bot }=c$, the Raleigh wave velocity. For a straight crack in a
system with uniform toughness, $\Gamma $, this leads to a monotonic
\begin{equation}
v(G^{\infty })=A^{-1}(G^{\infty }/\Gamma )
\end{equation}
for $G^{\infty }$ greater than the Griffith threshold, i.e. $G^{\infty
}>\Gamma $. When $G^{\infty }$ is smaller than the Griffith threshold
we assume that the crack does not move, i.e., once the solid breaks,
the crack does not to reheal (This is in fact observed in most
situations, with the absence of rehealing due to plastic and other
irreversible deformations at the crack tip). The velocity of the crack
is thus constrained to be positive.\cite{footnote2}

We are interested in the behaviour near the depinning transition at
which the crack starts to advance. We will use Eq.(\ref{generaleqn})
as the starting point of our analysis of the dynamics of the crack
front at this transition. The fracture toughness, $\Gamma $, in a {\it
heterogeneous solid} , is a position dependent quantity, which we
write as

\[
\Gamma (x,z)=\Gamma _{0}[1+\gamma (x,z)]
\]
with $\Gamma _{0}$ the mean value of the fracture toughness and
$\Gamma _{0}\gamma (x,z)$ the variable part of the fracture toughness
which we will take to have a zero mean and covariance given by
\begin{equation}
\left\langle \gamma (x,z)\gamma (x^{\prime },z^{\prime })\right\rangle
=\Upsilon (x-x^{\prime },z-z^{\prime })
\end{equation}
with a function $\Upsilon $ which is, generally, short-ranged in space.

The available energy, ${\cal G},$ is a complicated nonlinear
functional of the crack shape. In order to make progress, we will
expand the position of the crack front in powers of the deviation,
$f(z,t),$ away from a straight crack. The position of the crack-front
can be written as
\begin{equation}
F(z,t)=F_{I}+f(z,t),
\end{equation}
where $F_{I}$ is the original length of the crack which is assumed to
be much larger than the scales of motion of the crack front so that
the applied stress intensity factor, $K^{\infty },$ does not increase
significantly as the crack advances. Thus, the stored elastic energy
available to the crack front can thus be written in the form

\[
{\cal G}=G^{\infty }[1+g(z,t,\{f\})] 
\]
where $G^{\infty }$ is that for a straight crack of length $F_{I}$ for
the given external load. If $\partial f/\partial z$ is small, so will
be $g.$ To linear order in $f$, $g$ can be written as,

\[
g=-P\otimes f
\]
where $P$ is a kernel and $\otimes $ represents a convolution in space
and time. For a tensile crack, the Fourier transform of $P$ is
\cite{quasistatic} \cite{willis},

\begin{equation}
P(k,\omega )=\left\{ {}\right. 2\sqrt{k^{2}-\omega ^{2}/c^{2}}-\sqrt{
k^{2}-\omega ^{2}/a^{2}}+\frac{1}{2\pi i}\oint dW\tilde{I}(W,k^{2},\omega
^{2})\left. {}\right\}   \label{veckernel}
\end{equation}
with 
\begin{equation}
\tilde{I}=\frac{-\omega ^{2}}{\sqrt{Wk^{2}-\omega ^{2}}W^{3/2}}\ln \left[
\left( 2-\frac{W}{b^{2}}\right) ^{2}-4\sqrt{1-\frac{W}{a^{2}}}\sqrt{1-\frac{W
}{b^{2}}}\right] ;  \label{n7}
\end{equation}
the contour integral circling in the counter clock-wise direction the
cut in the complex $W$ plane that runs from $W=b^{2}$ to $W=a^{2}$
with $a$ and $b$ the longitudinal and transverse sound velocities
respectively; the argument of the logarithm in Eq.~(\ref{n7}) the
function whose zero, $W_{0},$ determines the Raleigh wave velocity via
$W_{0}=c^{2}$; and $\omega \to \omega +i0$ needed to define all cuts
e.g.,
\begin{equation}
\sqrt{k^{2}-\omega ^{2}/c^{2}}=-isign(\omega )|\sqrt{\omega 
^{2}/c^{2}-k^{2}}
|
\end{equation}
for $\omega ^{2}\geq c^{2}k^{2}$. Thus, 
\begin{equation}
P(k=0,\omega )=-i\omega B
\label{beq}
\end{equation}
where $B$ is a positive number.

To linear order in $f,$ the equation of motion can be written as

\begin{equation}
P\otimes f=-\gamma (z,t)+{\cal E}  \label{lineq}
\end{equation}
with the constraint that the local velocity of the crack front be positive
and 
\begin{equation}
{\cal E}=\frac{G^{\infty }-\Gamma _{0}}{G^{\infty }}
\end{equation}
which acts like the applied driving force on the crack front. From the
general structure of the energy release, ${\cal G}$, from
Eq.(\ref{gstructure}), we can separate the kernel $P$ into
sum of two terms, one which just depends only on the local velocity of
the crack front and the other which depends non-locally on the shape
of the crack front at all prior times.  Thus, we can express $P$ as
\begin{equation}
P(k,\omega )=-i\omega B+|k|\tilde{P}(\omega /|k|)  \label{eq18}
\end{equation}
where $|k|\tilde{P}(\omega /|k|)$ is the non-local part which vanishes as $
\omega \rightarrow \infty .$ The equation of motion can then be written in
the form

\begin{eqnarray}
B\partial _{t}f &=&\left\{ {\cal P}\int\limits_{z^{\prime },t^{\prime
}<t}J(z-z^{\prime },t-t^{\prime })\partial _{t^{\prime }}f(z^{\prime
},t^{\prime })-\gamma (z,t)+{\cal E}\right\} \times   \nonumber \\
&&\Theta \left[ {\cal P}\int\limits_{z^{\prime },t^{\prime }<t}J(z-z^{\prime
},t-t^{\prime })\partial _{t^{\prime }}f(z^{\prime },t^{\prime })-\gamma
(z,t)+{\cal E}\right]   \label{reapsp}
\end{eqnarray}
where 
\begin{equation}
J(z,t)=\int \int e^{ikz-i\omega t}[-\frac{|k|}{i\omega }\tilde{P}(\omega
/|k|)]
\end{equation}
and ${\cal P}$ denotes the principal part of the $z$ integral. The 
Heavyside
step function, $\Theta {\bf ,}$ constrains the velocity of the crack front
to be positive and will not be written out explicitly henceforth.

The kernel $J$ is readily evaluated from Eq.(\ref{veckernel}) to be 
\begin{eqnarray}
J(z,t) &=&-\frac{at\Theta (at-|z|)}{\pi z^{2}(a^{2}t^{2}-z^{2})^{1/2}}+\frac{
2ct\Theta (ct-|z|)}{\pi z^{2}(c^{2}t^{2}-z^{2})^{1/2}}  \nonumber \\
&&+\frac{1}{2\pi ^{2}i}\oint \frac{t}{\sqrt{W}(Wt^{2}-z^{2})^{3/2}}\ln
\left[ \left( 2-\frac{W}{b^{2}}\right) ^{2}-4\sqrt{1-\frac{W}{a^{2}}}\sqrt{1-
\frac{W}{b^{2}}}\right]   \label{jvec}
\end{eqnarray}
where the branch cuts are defined as previously.

The stress transfer kernel Eq.(\ref{jvec}) is rather complicated. It
is plotted as a function of $t$ for a fixed $z$ in Fig[\ref{fig:jvec}]
for a Poissions ratio, $\nu =0.25$. At the arrival time of the Raleigh
waves, $J$ diverges as $1/(ct-z)^{1/2}$ and then decays slowly to its
long time value, i.e., $J(z,t\rightarrow\infty)\rightarrow
\frac{1}{\pi z^{2}}$, the static stress transfer kernel. Although the
negative stress precursor to the stress peak that occurs for
$z/a<t<z/c$ may well be important, for our primary purposes here, we
believe that the stress peak is the more important feature. It is
therefore useful to study a somewhat simpler model which has a similar
stress peak.

We choose to study a {\it scalar approximation} to elasticity
theory. In this approximation the displacement field in the solid is
take to be a scalar field $\phi ,$ satisfying the three dimensional
scalar wave equation

\begin{equation}
\frac{1}{c^{2}}\partial _{t}^{2}\varphi -\nabla ^{2}\varphi =0
\label{scalarwaveeq}
\end{equation}
The displacement field $\varphi $ has a discontinuity across the crack
surface while the normal derivative $\partial _{y}\varphi (y=0^{\pm
})$ (the ``stress'')\ vanishes on the crack surface. We shall refer to
this model as the {\it scalar model}. Under the external load, the
displacement field $ \varphi $ has a $\sqrt{F(z,t)-x}$ singularity at
the crack front proportional to the scalar stress intensity factor
$K(z,t)$ as for real elasticity. The corresponding kernel $P$ can be
written in Fourier space as \cite{perrin}

\begin{equation}
P_{{\rm {scalar}}}=\sqrt{k^{2}-\omega ^{2}/c^{2}}  \label{4scalarkernel}
\end{equation}
and the stress transfer kernel is 
\begin{equation}
J_{{\rm se}}(z,t)=\frac{ct\Theta (ct-|z|)}{\pi z^{2}(c^{2}t^{2}-z^{2})^{1/2}}
\label{44scalarkernel}
\end{equation}
We see that the stress peak, the long time tail and the static stress
transfer kernel $J(z,t\rightarrow \infty )$ are all of similar form to
the real tensile crack case.

From the kernels Eq.(\ref{jvec}) and Eq.(\ref{44scalarkernel}) we see
that for both the tensile crack and the scalar model, sound waves
yield non-monotonic kernels which lead, in response to a jump of one
segment of the crack front, to ephemeral overshoots of the stress
above the eventual static value. The magnitude of the overshoots will
be governed by microscopic factors such as the microscopic response
time of the crack front and acoustic damping processes which can be
incorporated by the replacement of $\omega ^{2}$ by $ \Omega^{2}
=\frac{\omega^{2}}{1-i\omega \tau _{d}}$ in Eq.(\ref{veckernel}) or Eq.(\ref
{4scalarkernel}) with $\tau _{d}$ an acoustic relaxation time. We are
interested in how these overshoots affect the dynamics of the crack
front near threshold. But in the limit that all crack disturbances
move along the crack slowly compared to $c$, we can neglect the
effects of sound waves, and the transfer of stress will be effectively
instantaneous yielding the quasistatic model with the kernel given by
\cite{ricelongrange}
\begin{equation}
J_{{\rm {qs}}}(z,t)=\frac{1}{\pi z^{2}}\Theta (t).  \label{jqs}
\end{equation}
This quasistatic model we study first, its possible regimes of validity are
discussed in Section V.

In order to separate the effects of sound travel time delays from
those of stress pulses, we also consider a model with monotonic stress
transfer characterized by the kernel
\begin{equation}
J_{{\rm {td}}}(z,t)=\frac{1}{\pi z^{2}}\Theta (t-|z|).
\label{jtd}
\end{equation}
This kernel is similar to the quasistatic kernel except that the
stress transfer is not instantaneous. Finally, in order to separate
the effects of the maximum of the stress peaks from those of their
tails, and to make the analysis of the stress peaks more tractable, we
study a kernel with sharp pulses defined in Eq.(\ref{sharppulsekernel}). 
In both of these
artificial models, the velocity of signal propagation has been set
equal to one.  Snapshots of the stress pulses when the crack front at
$z=0$ is moved ahead by a small amount at $t=0$ and held there, are
shown for the various models in Fig[\ref{pulseplot}].  These are
just plots of the respective $J(z,t)$ for a fixed $t.$

In general, the stress transfer along the crack front will depend on
the shape of the front due to non-linear terms in the expansion of
${\cal G}[\{f\}]$ in powers of $f$. Throughout this paper we will
ignore these. We can justify this approximation for the quasistatic
case for which we have an analytic understanding, and believe that it
should generally be valid on long length and time scales as long as
the crack front roughness exponent $\zeta <1$; i.e., that the crack
front looks straight on asymptotically long scales.\cite{quasistatic}

\subsection{Numerical Implementation}

\label{numimp} 
We are interested in the behaviour of the crack front near to when it
begins to move. Below and just above the critical load, the motion of
the crack front is very jerky and segments of the crack front move
ahead and then get stuck in a tougher region. The basic minimum
length, time and increment in the crack front position scales of these
processes are set by the length scales of the toughness variations,
the coefficient $B$ in Eq.(\ref{beq}) etc. Thus to understand this
behaviour, it is natural to simulate the crack front in a manner which
is discrete in space, time and position.

Our simulations were done on a lattice which is periodic in the
direction, $x$, of crack advance, but for each $z=0...L-1,$ the
co-ordinate along the crack front, the row of points are shifted by an
independent random amount, $b(z)$ with $0<b(z)<1.$ The allowed values
of the crack front position are thus
\begin{equation}
f(z)=b(z)+n(z)
\end{equation}
with $n(z)$ integers. This avoids the possibility of phase locked
behaviour in which the crack advances in a relatively uniform manner
characterized by all points advancing by one before any advance
again. Periodic boundary conditions are used in the $z-$direction. We
have chosen the sound velocity in our models to be unity corresponding
to one lattice spacing along the crack front per time step (all the
models we study numerically have only one sound velocity).

At each lattice point an independent value of the random fracture
toughness, $\gamma (x,z),$ is picked from the interval [0, 1.5]. This
range is chosen so that the variations in $\gamma$ are comparable to
the force on points of the crack front on each other which are
\begin{equation}
g(z,t)=\sum\limits_{z'=0}^{L-1}\sum\limits_{t'\leq
t}\tilde{J}(||z-z'||,t-t')[f(z',t')-f(z',t'-1)]
\label{g1}
\end{equation}
where, with the periodic boundary condition in $z$ on a crack of
length $L$;
\begin{equation}
||z-z'||\equiv min(|z-z'|,|L-|z-z'||),
\end{equation}
 is the shortest distance between $z$ and $z'$. The stress transfer
kernels $\tilde{J}$ are modifications of the continuum kernels of
interest with the stress pulses designed to die away smoothly after
going through the system once. Thus, although there is a long ranged
history dependence in all but the quasistatic model, we need keep
track of the history of the interface only up to a time corresponding
to the sound travel time through half the system, i.e., a time of
$L/2,$ where $L$ is the system size.  Thus,
\begin{equation}
\tilde{J}(z,t\geq L/2)=\tilde{J}(z,\infty)=\tilde{J}_{qs}(z)=\frac{1}{||z||^2}.
\label{jqsdis}
\end{equation}
for $z\neq 0$. The sum in Eq.(\ref{g1}) over $-\infty\leq t'\leq
t-L/2-1$ can thus be replaced by
$\sum\limits_{z'}\tilde{J}(||z-z'||)f(z',t-L/2-1)$.  Care must
also be taken with the ``self-interaction'' piece $z=z'$ which
represents the ``principal part'' in Eq.(\ref{reapsp}). In the
quasistatic case (and generally for long time)
\begin{equation}
\tilde{J}(z=0,t)=-\sum\limits_{z\neq 0}\tilde{J}_{qs}(||z||)
\label{principal}
\end{equation}
so that if the crack moves uniformly there are no changes in $g$. More
 generally, in particular for the artificial models,
 $\tilde{J}(z=0,t)$ involves some arbitrariness. To preserve the
 monotonicity of the sound-travel-time delayed model, a certain choice
 is required. This, and the detailed form of the various
 $\tilde{J}(z,t)$ used are specified in Appendix \ref{kernelsnum}. 
The evaluation of the elastic force at each time step is done in
Fourier space using the FFT algorithm and hence the time for
computation at each time step of the evolution of the interface scales
like $L^{2}\log L$.

The ``driving
 force'', ${\cal E},$ is the forcing parameter and as this is
 increased the crack begins to move. When the total ``force'' at a
 point $z$ on the crack front is greater than the random part of the
 fracture toughness there, the crack front at $z$ advances by one
 lattice constant, i.e.,

\begin{equation}
f(z,t+1)=f(z,t)+{\Theta }{\huge [}g(z,t)-\gamma (z,t)
+{\cal E}{\huge ]}  \label{discreteeqn}
\end{equation}
where the lattice constant is set equal to one and ${\Theta }$ is the
step function.

These discrete automaton models for the crack front are expected to
capture the physics at threshold of the corresponding continuum
models at length scales long compared to the correlation length of
the random toughness.  Direct evidence for universality is provided by
extensive numerical simulations on charge density wave
models\cite{am1} in two dimensions which have found universal
behaviour for smooth and piecewise continuous pinning forces as well
as for discrete cellular automata analogous to the one defined above. We
expect the same to hold here.

\section{Monotonic Models}

\label{monor}

\subsection{Quasistatic Model}
\label{quasistaticr} 

We first consider the quasistatic model.  In this approximation, the
 stress transfer is instantaneous and the linearized continuum{\it \
 }equation of motion of the crackfront takes the form

\begin{equation}
\partial _{t}f(z,t)=\frac{1}{\pi }{\cal P}\int dz^{\prime }\frac{f(z^{\prime
},t)-f(z,t)}{(z-z^{\prime })^{2}}-\gamma [f(z,t),z]+{\cal E}  \label{4modq}
\end{equation}

Before presenting the numerical results from which we determine the
values of various critical exponents characterizing the depinning
transition, we give, following references \cite{on,de}, scaling
arguments for several identities between the exponents and bounds on
them.

\subsubsection{Exponent Identities and Bounds}

In the moving phase, the crackfront will be reasonably smooth at
scales larger than the correlation length $\xi ,$ indeed at large
scales $f\approx vt$ and hence the random toughness, $\gamma (f,z)$,
will act essentially like white noise and hence $\langle
(f(z,t)-f(0,t))^{2}\rangle \sim \ln (z).$ On scales smaller than $\xi
,$ the front will be rough with $|f(z,t)-f(0,t)|\sim |z|^{\zeta }.$ 

In order for the crack motion to be smooth on larger scales, each segment
of the crack of length $\xi $ must take about the same time $\tau \sim
\xi ^{z}$ the correlation time, to move through each distance $\xi
^{\zeta }.$
In the region which a segment of length $\xi$ passes through in time
$\tau$, there are $\xi ^{\zeta +1}$ random values of the local
toughness. This means that the force per unit length needed to pull
the crack segment through this region must vary from region to region
by at least of order the random variation in the toughness averaged
over this region, i.e., $1/\sqrt{\xi ^{\zeta +1}}$ by the central
limit theorem.  Thus the course grained toughness variations at the
scale $\xi $ are
\begin{equation}
\delta \Gamma _{\xi }\gtrsim 1/\xi ^{(\zeta +1)/2}.
\end{equation}
The force on the segment from the external load and the rest of the
crack must be just strong enough to overcome these random
variations. If these forces were too strong, the segment would move
more smoothly implying that it must have been longer than $\xi $ by
definition. On the other hand, if they were too weak, the segment
would not move at all in some time intervals of length $\tau $ and
thus it must have been smaller than $\xi .$ Since the mean external
load at which the segment moves is
\begin{equation}
G_{c}\equiv G_{c}^{qs},
\end{equation}
this implies that either 
\begin{equation}
G^{\infty }-G_{c}\sim \delta \Gamma _{\xi }\gtrsim 1/\xi ^{(\zeta +1)/2}
\end{equation}
i.e., 
\begin{equation}
\nu \geq \frac{2}{\zeta +1},  \label{2.39}
\end{equation}
or the force per unit length from the neighboring sections of the
crack, $ G_{n}$, are comparable to $\delta \Gamma _{\xi }.$ The latter
is dominated by nearby segments so that
\begin{equation}
G_{n}\sim \int\limits_{\xi }^{2\xi }dz\frac{z^{\zeta }}{z^{2}}\sim \xi
^{\zeta -1},
\end{equation}
yielding 
\begin{equation}
\xi ^{\zeta -1}\geq \frac{1}{\xi ^{(\zeta +1)/2}}
\end{equation}
and hence, 
\begin{equation}
\zeta \geq 1/3.
\end{equation}
If there is only one basic scale of the forces near threshold, as
simple scaling would suggest, then we should expect that
\begin{equation}
G^{\infty}-G_{c}\sim G_{n}\sim \delta \Gamma _{\xi }.
\end{equation}
The rough equality of the typical force per unit length, $G_{m}$,
 of a segment
of length $\xi $ on a segment a distance $\xi $ away and
$G^{\infty}-G_{c}$ thereby yields the scaling relation
\begin{equation}
\nu =\frac{1}{1-\zeta }.  \label{nuze}
\end{equation}
Similar argument can be used below threshold implying that the
correlation length exponents on the two sides of the transition are
equal\cite{on}.  We will derive the relation Eq.(\ref{nuze})
 more directly below.

The bound on $\nu,$ Eq.(\ref{2.39}), comes from an argument similar to
that by Harris\cite{harris} for equilibrium phase transitions and
established more generally in reference \cite{ccfs}$.$ We can also
derive it by considering some segment of the crack front of size
$\xi_{-} $ below the threshold, loosely defining $ G_{\xi
}^{c}=G_{c}+\delta \Gamma _{\xi }$ to be the local critical force at
which this segment begins to move. Below threshold, segments of size
of order $\xi $ have a substantial chance both of having already moved
or not having moved yet while $G^{\infty }$ was increased to its
present value. Thus the variations in $G_{\xi }^{c}$ must be
comparable to $G_{c}-G^{\infty }.$ This is similar to what was argued
for above $ G_{c}, $ but here it does not rely on as many assumptions
about scaling.  Since $\delta \Gamma _{\xi }\gtrsim 1/\xi_{-}
^{(1+\zeta )/2},$ we obtain Eq.(\ref{2.39}) just as from above
threshold.

We now obtain scaling relations for the distribution of avalanche
sizes below the threshold loading. Following Ref.\cite{on} we
conjecture that the distribution of avalanches as $G^{\infty}$ is
increased slightly has a scaling form
\begin{equation}
{\rm Fraction}{\rm\ of\ avalanches\ with\ size \mbox{$>$}}l\ {\rm when}
 \ G^{\infty}\rightarrow G^{\infty}+dG^{\infty}\approx \frac{1}{l^{\kappa
}}\hat{\rho}(l/\xi _{-})
\end{equation}
at a given external load, with $\xi _{-}\sim (G_{c}-G^{\infty })^{-\nu }.$
Following the same reference we obtain, 
\begin{equation}
\kappa =1-1/\nu  \label{kappa}
\end{equation}
from the increase in the mean position as $G_{c}$ is approached.  Now
consider the probability distribution of the size of {\it all}
avalanches that occur on sweeping the load from zero to the threshold
load,
\begin{equation}
{\rm {Fraction\ }}{\rm {of\ all\ avalanches\ with\ size\  \mbox{$>$}}}l
\approx
\int\limits_{0}^{G_{c}}\frac{1}{l^{\kappa }}\hat{\rho}(l/\xi _{-})n_{A}
(G^{\infty})dG^{\infty
}\sim l^{-\kappa -1/\nu }=l^{-1}
\end{equation}
where the last equality was obtained using the scaling relation
Eq.(\ref {kappa}) and the observation that the rate of avalanche
production, $n_{A}(G^{\infty})$, 
per increase in $G^{\infty}$ goes to a {\it constant}
at $G_{c}$. Thus,
\begin{equation}
{\rm {Prob}}({\rm {size\ of\ a\ given\ avalanche =}}l)
\sim 1/l^{2}  \label{avdist}
\end{equation}

An exponent identity relating the velocity exponent, $\beta ,$ to the
other exponents follows directly from the picture discussed above of
the moving segments of length $\xi .$ Since the time for a segment of
length $\xi $ to move ahead a distance which scales as $\xi ^{\zeta
}$, is of order $\xi ^{z},$ the velocity of the interface scales as
$\xi ^{\zeta -z}.$ We thus obtain
\begin{equation}
\beta =(z-\zeta )\nu .  \label{betaexp}
\end{equation}

Another useful relation can be obtained by considering adding an
additional ``force'', ${\epsilon} (z,t)$, on the crack front. Denoting
the resulting change $\delta f(z,t),$ we can define the polarizability
$\chi $ as
\begin{equation}
\chi (k,\omega )=\frac{\delta \langle f(k,\omega )\rangle }{\delta {\epsilon}
(k,\omega )}  \label{245}
\end{equation}
in Fourier space. First consider applying a static force. In this case
the additional force ${\epsilon} (k,\omega )$ can be absorbed by
redefining
\begin{equation}
f(k,\omega )\rightarrow f^{\prime }(k,\omega )-\frac{{\epsilon} (k)}{|k|},
\end{equation}
since the terms from the interaction of the crack front with itself in
the equation of motion will then exactly cancel the additional force
${\epsilon}.$ The statistics of the random toughness variables, in
this distorted frame,
\begin{equation}
\gamma ^{\prime }(x,z)=\gamma (x+\phi (z),z),
\end{equation}
with $\phi (z)$ the Fourier transform of ${\epsilon} (k)/|k|,$ will
have the {\it same statistics} as the original ones.  This is an
important {\it statistical symmetry }of the system. It is the small
angle form of the statistical rotational invariance. We thus have $
\delta \langle f(k,\omega )\rangle =\frac{{\epsilon} (k)}{|k|},$ and
hence
\begin{equation}
\chi (k,0)=\frac{1}{|k|}  \label{chistatic}
\end{equation}
{\it exactly}. On the other hand, on applying a low frequency 
spatially uniform force, ${
\epsilon} (\omega ),$ we should have 
\begin{equation}
-i\omega \chi (0,\omega )=\frac{dv}{d{\epsilon} }\sim (G^{\infty
}-G_{c})^{\beta -1}.  \label{eq35}
\end{equation}
Generally, we expect that $\chi (k,\omega )$ will have the scaling
form
\begin{equation}
\chi (k,\omega )\sim \frac{(G^{\infty }-G_{c})^{\beta -1}}{-i\omega }X(k\xi
,\omega \xi ^{z})  \label{eq36}
\end{equation}
with the form of the prefactors implied by Eq.(\ref{eq35}). In
the static limit, ${X(k\xi, u=0)}=0$, and 
\begin{equation}
\lim\limits_{\omega \rightarrow 0}\chi (k,\omega )\sim \xi ^{z}(G^{\infty
}-G_{c})^{\beta -1}i\frac{\partial X(k\xi, u)}{\partial u}|_{u=0}.  
\label{chilim}
\end{equation}
Comparing Eq.(\ref{chistatic}) and Eq.(\ref{chilim}) we see that 
\begin{equation}
\frac{\xi ^{z}(G^{\infty }-G_{c})^{\beta -1}}{|k|\xi }\sim \frac{1}{|k|}.
\end{equation}
But since $\xi \sim (G^{\infty }-G_{c})^{-\nu },$ we have 
\begin{equation}
\beta -1-\nu (z-1)=0.  \label{253}
\end{equation}
Using the expression for $\beta $ from Eq.(\ref{betaexp}) we again
obtain Eq.(\ref{nuze}). As noted above, this simply relates the force
of length $ \xi $ segments on each other to $G^{\infty }-G_{c}.$

We thus have two independent exponents from which the others can be
obtained. In addition, from Eq.(\ref{2.39}) and Eq.(\ref{betaexp}) we
obtain the bounds
\begin{equation}
\zeta \geq 1/3.  \label{zetabound}
\end{equation}
and hence,
\begin{equation}
\nu \geq 3/2
\end{equation}
All of the exponent identities and the form of scaling functions such
as Eq.(\ref{eq36}) have been derived from a renormalization group
expansion about two dimensions which is the critical dimension for the
depinning transition of manifolds driven through random media with
long range interactions decaying as $1/r^{d+1},$ --- i.e., $|k|$ in
Fourier space\cite{CL}. The analytical results from the
$d=2-\varepsilon $ expansion are compared with our numerical results
in the next subsection.

\subsubsection{Numerical Results}

We now present the numerical results for the quasistatic model from
which we obtain the values of the various exponents. As discussed
earlier, we simulate a discretized version of this equation, here
simply
\begin{equation}
f(z,t+1)=f(z,t)+{\bf {\Theta }}[\sum\limits_{z=0}^{L-1}
\frac{f(z^{\prime },t)-f(z,t)}{||z-z^{\prime }||^{2}}-\gamma (z,t)+{\cal E}]
\end{equation}
where $L$ is the system size, to study the dynamics at threshold. We
start with a pinned configuration of the crack front that is as close
as possible to being straight and gradually increase the load until
the most weakly pinned point becomes unstable and jumps. This point,
in turn, may pull along other points on the crack front due to the
elastic interactions, causing an avalanche. During the avalanche, the
load is held fixed. Once the avalanche subsides, the load is increased
once again until another point becomes unstable, and so on.  A series
of avalanches, as the load is gradually increased in this way is shown
in Fig[\ref{avmodq}].  A space-time plot of one of the large
avalanches is shown in Fig[\ref{avpic} ].

Defining the {\it size} of an avalanche is somewhat problematic. We
have chosen to define it as the number of distinct points on the crack
front that move during the course of the avalanche. Note that various 
other ways of
defining the ``size'' of an avalanche along the crack front by e.g.,
its ``moment of inertia'' about its center of mass or by its maximum
extent have problems because of the power law tail of the interactions
which can trigger some jumps far away. In addition, periodic boundary
conditions would complicate a definition. To study the statistics of
many avalanches, the avalanche sizes are binned in powers of two. To
measure the dynamic exponent $z$, statistics of avalanche sizes versus
their durations are collected for all the avalanches that occur
as the load is increased from zero to the critical load. The log-log
plot of the number of avalanches in a given bin against the bin size is
shown in Fig.[\ref {fig:binsize}] and a linear fit gives us a slope of
$2.14\pm 0.3$ in agreement with Eq.(\ref{avdist}) but with large
errors.

Figure[\ref{zexp}] shows the plot of the mean duration of avalanches
in a bin, $\tau_{bin}$ versus bin size, $l_{bin}$, for a system of
length 1024.  From the slope of the log-log plot we determine
\begin{equation}
z=0.74\pm 0.03.
\end{equation}
As is generally true, one must be very careful not to take such
statistical estimates of uncertainties in exponents at face value due
to the existence of corrections to scaling. Fortunately, in our case,
the 2$-\varepsilon $ expansion provides an estimate of the leading
correction to scaling exponent; calculations of
Narayan\cite{onprivate} yield the leading irrelevant eigenvalue at the
critical fixed point to be approximately $-\varepsilon /3\approx -1/3$
in our case with $\varepsilon =1.$ We thus fit the data to the form
$\tau_{bin}=\frac{Cl_{{\rm bin}}^{z}}{1+A_{z}l_{{\rm bin}}^{-1/3}}$
and find that $A_{z}<<1$ and hence this fit gives the same value of
$z$ to within error bars.

From the $\varepsilon -$expansion\cite{on,de}, it was found that
\begin{equation}
z\approx 1-2\varepsilon /9+{\cal O(}\varepsilon ^{2})\approx 7/9\approx 
0.78
\end{equation}
for $\varepsilon =2-d=1.$ If we neglect the ${\cal O(}\varepsilon
^{2})$ and higher terms, this agrees with our numerical result within
error bars.

In a finite system, there is some ambiguity in the definition of the
critical load. For example, if the system extends very far in the
direction of motion, the whole crack front would typically move from
its initial position but get stuck in a rare tough region far away.
In order not to bias the results by choice of the system extent in the
direction of motion, we define the critical load as the load at which
every point but one on the interface has moved at least once.  For a
large system we find
\begin{equation}
G_{c}\approx 0.97.
\end{equation}
Note that the random forces, the critical driving force and the nearest 
neighbor
elastic forces are very comparable.

Right at threshold the crack front
is found to be self-affine as expected. Figure[\ref{zeta}] shows the
plot of the power spectrum of the crack front, $\langle
|f(k)|^{2}\rangle $, at threshold, as a function of the wavevector for
various system sizes ranging from 4 to 4096.  We expect the power 
spectrum to be $k^{-(2\zeta +1)}$ for small $k$. The best fit to a
straight line is shown in Fig[\ref{zetaslope}] for a system size of
4096 averaged over 1000 samples. The slope of this line gives us $
2\zeta +1$, from which we determine
\begin{equation}
\zeta =0.34\pm 0.02.  \label{259}
\end{equation}
Surprisingly, even at very small wavevectors, the power spectrum still
looks linear on a log-log plot and we do not see significant finite
size effects even at the wavevector corresponding to half the system
size. Since the data in Fig[\ref{zetaslope}] have very small
statistical uncertainties, we can try to fit the power spectrum to the
form $Ck^{-(2\zeta +1)}(1+A_{\zeta}k^{1/3})$ using corrections to scaling. 
The
coefficient $A_{\zeta}$ turns out to be very small and we obtain the same
roughness exponent with comparable error bars. This gives us some
confidence in the estimate Eq.(\ref{259}).

Our result for $\zeta$ satisfies --- and may saturate--- the bound
$\zeta \geq 1/3.$ The prediction of $\zeta $ from the $\varepsilon $
expansion is
\begin{equation}
\zeta =\varepsilon /3+{\ o}(\varepsilon ^{n})  \label{260}
\end{equation}
for all $n$\cite{on}; i.e., there appears to be no corrections to all
orders in $ \varepsilon $ although ``non-perturbative'' corrections
cannot be ruled out.  Nevertheless, the bound Eq.(\ref{zetabound}),
the $\varepsilon $ expansion result Eq.(\ref{260}), and the numerics
suggest that perhaps $\zeta $ may be exactly $1/3$ although at this
point we know of no solid argument that yields 1/3 as an upper (to
complement the lower) bound.

The correlation exponent, $\nu $, can be obtained via the finite size
scaling hypothesis by measuring the variance of the threshold load,
$G_{c}(L)$, as a function of the system size. 
Assuming that there is only one important
length scale $\xi ,$ the variance of the threshold force, $(\Delta
G_{c}(L))^{2}$ scales with the system length, $L$, as 
\begin{equation}
(\Delta G_{c}(L))^{2}\sim L^{-2/\nu }.
\end{equation}

A direct fit to $L^{-1/\nu }$ of the plot in Fig[\ref{nulinearfit}] of
the variance of the threshold load versus the system size, for system
lengths ranging from 4 to 8192, leads to $\nu =1.80\pm 0.05,$ while a
fit for system lengths ranging from 256 to 8192, gives
$\nu=1.72\pm0.12$.  But a systematic curvature can be seen. In light
of the knowledge of the corrections to scaling, we can do better by
fitting to the form $\frac{ CL^{-1/\nu }}{1+A_{\nu}L^{-1/3}}$ from which
obtain
\begin{equation}
\nu =1.52\pm 0.02.
\end{equation}
This fit is shown in Figure[\ref{correctoscalingfit}] for systems of
length 4 to 8192 lattice constants. In this case, as suggested by the
data, $A_{\nu}$ is not small and the fit indeed yields $A_{\nu}=0.87$. 
Note that
the expected scaling equality Eq.(\ref{nuze}) is obeyed, but {\it only
when the corrections to scaling are included}.

As the load is increased above threshold, the crack front begins to
move. As for the critical load, we must be careful how we define the
velocity for finite length crack fronts. If we choose a system of
great extent, $W,$ in the direction of motion, the front will tend to
get stuck in anomalously tough regions; this effect will be more
pronounced for small $L$. But since we are interested in the critical
behaviour and we have a good handle on the scaling of $f$ with $L$, we
can instead choose systems of extent $W\approx C_{W}L^{\zeta }$ with
periodic boundary conditions in the direction of motion.  For
monotonic models, the convexity then implies convergence to a unique
steady state\cite{nopassing}. Since at threshold $\Delta f\sim
L^{\zeta }$ with a coefficient roughly of order unity, we choose
$C_{W}=4.$

There is a complication that must be considered: due to the
 possibility of a pinned configuration for loads above that at which
 the last point became depinned, the minimum $G^{\infty }$ at which
 $v>0$ will sometimes be greater than our definition of $G_{c}$ by a
 random amount whose distribution depends on $C_{W}$.  From scaling we
 expect
\begin{equation}
G_{{\rm {min}}}^{{\rm {moving}}}-G_{c}\sim 1/L^{1/\nu }.
\end{equation}
By scaling, there will thus be a typical minimum velocity 
\begin{equation}
v_{{\text{min}} }\sim 1/L^{\beta /\nu }\sim 1/L^{z-\zeta }.
\end{equation}
Note that the minimum velocity due to the discreteness of time is much
less than this and hence negligible for large $L$.

Figure [\ref{beta}] shows a plot of the mean velocity of the front as
a function of the loading for a system of length 1024 from which we
determine
\begin{equation}
\beta =0.68\pm 0.06.
\end{equation}
The fit using the corrections to scaling leads to the same value of $\beta$
within the error bars. Surprisingly, there do not seem to be substantial 
deviations for $G\geq G^{{{\text{moving}}}}_{{\text{min}}}$.

The $\varepsilon -$expansion prediction is $\beta \simeq 2/3$ if we
use $\zeta =1/3$ and $z=7/9$; again there is reasonably good agreement
between our numerical results and the $ \varepsilon$ expansion although 
our error bars are larger for $\beta$ than for $z$, $\nu$ or
$\zeta$.

In our numerical results, the critical
force and all of the coefficients in scaling laws are of order unity
suggesting that we have no intermediate length, displacement or time
scales and thus that the scaling should work well for all but small
size samples. If we had chosen too narrow a distribution of random
toughness $\gamma(x,z)$, there would have been an intermediate length
scale and this would no longer have been the case.
 Note that the renormalization group methods can be used to show that
non-linearities associated with higher order terms in the expansion of
${\cal G}[\{f\}]$ are, for quasistatic stress transfer, irrelevant
for the critical behaviour. One could have guessed this since, from
the homogeneity of ${\cal G}[\{f\}]$ higher powers of $f$ have an
equivalent number of powers of gradients, so that $\zeta <1$ implies
that they are irrelevant.  
\subsection{ Monotonic model with Time Delayed Interaction.}

\label{timedelayr} 
In the previous section, we saw that the quasistatic approximation to
stress transfer gave rise to a critical depinning transition with a
dynamic exponent $z<1.$ This means that for large enough avalanches
which typically occur only if the load is close enough to $G_{c},$ the
effective disturbance velocity of an avalanche of size $l$ will be
$l/l^{z}$ times the basic microscopic velocity scale of disturbances
set by the dissipative coefficient $B$ in Eq.(\ref{eq18}). Thus for
sufficiently large $ l$, the quasistatic avalanches will progress
faster than the sound speed.  This is clearly unphysical and in this
regime, the sound travel-time delays in the stress transfer must play
a role. In order to understand the effects of these and of stress
overshoots separately, we study a monotonic model --- i.e. with no
stress overshoots --- but with sound travel-time delays. The simplest 
form
of this is to simply replace the $\Theta (z)$ in the quasistatic
stress transfer with $\Theta (t-|z|).$ On a lattice the stress
transfer kernel becomes:
\begin{equation}
J_{{\rm {td}}}(z,t)=\frac{\Theta (t-||z||)}{ ||z||^{2}}(1-\delta
_{z,0})-\delta _{z,0}\sum\limits_{z^{\prime}\neq 0}\frac{1}{
||z||^{\prime 2}} \label{delayedmodq}
\end{equation}
Some care is needed in choosing the second, local, term in Eq.(\ref
{delayedmodq}). The natural choice would be to fix $J(z=0,t)$ by the
condition that for all times, $\sum\limits_{z}J(z,t)=0.$ This
condition is satisfied for the quasistatic model and for the scalar
model as well as for real elasticity. It ensures that for a straight
crack, the instantaneous crack front velocity is a function solely of
the instantaneous external load and independent of the past history of
the crack, since for a straight crack the effect of the crack front
interactions vanish at all times. If we made $
\sum\limits_{z}J(z,t)=$0 here also, however, the model
Eq.(\ref{delayedmodq}) would no longer be monotonic. Rather, the
stress at a point $z,$ after a jump at the same point, would decrease
in time as the integrated stress transferred to the rest of the crack
increases; this would act like a stress overshoot.  For now we will,
therefore, give up the independence of a straight crack on its past
history to preserve the monotonicity condition.

We see that in this model the force on any given point on the crack
front at any given time is always less than or equal to the
equivalent force for the same configuration in the quasistatic
model. Note also that for a crack front which is stationary after some
time $t$, in the discrete-time periodic boundary-condition
version of this model (see Appendix \ref{kernelsnum}), the force at all points on the crack
front, will reach the quasistatic value by time $ t+L/2$. As shown in
Appendix \ref{nopass}, these conditions imply that if the external load is
increased adiabatically from the same initial conditions, the
time-delayed model, Eq.(\ref{delayedmodq}), has {\it exactly the same
static properties} as the quasistatic model. Indeed, for a given
realization of the random toughness, every finite avalanche that
occurs in the two models will be identical, except for the times at
which points on the crack front jump. Therefore, both the models will
have identical threshold forces, and the exponents $\nu $ and $\zeta $
are then obviously identical to their quasistatic values.

The space--time plot of a particular avalanche as one site is triggered by
increasing the load, is shown in Fig.[\ref{avmodqcomp}] for the
quasistatic model and of the identical avalanche in 
the sound-travel time delayed model starting
from the same initial configuration of the crackfront in a system of
length 512. It is evident that the dynamics of the avalanches in 
the two models is
very different

 The Fourier transform of the continuum version of the time-delayed
kernel in Eq.(\ref{delayedmodq}) is given by
\begin{equation}
J(k,\omega )=-\frac{1}{i\omega }\{-\frac{i}{\pi }(k+\omega )\ln |k+\omega |+
\frac{i}{\pi }(k-\omega )\ln |k-\omega |-\frac{1}{2}|k+\omega |-\frac{1}{2}
|k-\omega |\}+U(\omega )
\end{equation}
where there is an ambiguity in the uniform $k=0$ part of the Fourier
transform, $U(\omega )$. It is clear that the dynamic exponent must be $
z\geq 1.$ Let us assume that $z>1$ or more precisely that the 
characteristic
time $\tau \gg \xi $ near threshold. We are thus interested in the behaviour
in the scaling limit in which  $|\omega |\ll |k|.$ In this limit 
\begin{equation}
J(k,\omega )\approx \frac{2}{\pi }\ln |k|+\frac{|k|}{i\omega }
\end{equation}
From the equation of motion, the response function 
defined in Eq.(\ref{245}) is in the absence of the randomness 
\begin{equation}
\chi =\frac{1}{-i\omega B+i\omega J(k,\omega )}\approx \frac{1}{-i\omega 
\frac{2}{\pi }\ln (\frac{1}{|k|})+|k|}
\end{equation}
for $\omega \ll k,$ since the ln$|k|$ term from $J$ will dominate the $
\omega $ dependence. In mean field theory, this gives rise to times 
scaling
with lengths as 
\begin{equation}
\tau \sim \xi \ln \xi .
\end{equation}
In the absence of the ln
\mbox{$\vert$}%
$k|$, renormalization due to the random roughness would make $B_{{\rm
{eff}} } $ decrease with length scale in $2-\varepsilon $
dimensions. However, the ln$|k|$, being singular, cannot
renormalize. But it {\it can} feed into the renormalization of $B.$
Following this through yields
\begin{equation}
\tau _{\xi }\sim \tilde{c}\xi
\end{equation}
in dimensions $d=2-\varepsilon $ with $\tilde{c}$ an effective velocity of
order $\varepsilon $ for small $\varepsilon ,$ but presumably of order 
unity
in our one-dimensional case. Thus we see that assuming $z>1$ leads 
back to 
\begin{equation}
z=1
\end{equation}
which we believe should be correct with {\it no }logarithmic
corrections.  Using the exponent identities Eq.(\ref{betaexp}) and
Eq.(\ref{nuze}) we obtain
\begin{equation}
\beta =1.
\end{equation}
A plot of the load versus the mean crack velocity is shown in Fig[\ref
{delayedmodqbeta}] for a system of size 64. It is very close to linear
although the range is small enough that one cannot reliably extract
$\beta$. Note that because of the dependence on the past history, we
are limited here to rather small samples. While the
computations for each time step for the quasistatic model take a time
of order $L{\rm ln}L$, those for the sound-travel time delayed model
take $L^{2}{\rm ln}L$ and hence even for a system of length 64, the
statistics are more difficult to obtain.

\section{Stress Overshoots: Non-Monotonic Kernels}

\label{nonmonor} 
We now turn to more realistic stress transfer kernels. We have seen,
from Eq.[\ref{4scalarkernel}] and Eq.[\ref{veckernel}], that bulk
sound modes naturally lead to non-monotonic kernels of the stress
transfer along the crack front. A regime of negative stress transfer,
as occurs at intermediate times for tensile cracks, cannot, by itself,
change the behaviour much from the time delayed monotonic
models since, in the absence of stress overshoots, the static
behaviour will again be identical to the quasistatic model. Thus the
primary differences between the time-delayed model and more realistic
models must be associated with the stress overshoots. The actual shape
of the stress overshoot may be complicated by various factors
including the effects of multiple scattering of sound waves off the
crack front. Therefore, we would like to understand what features of
the stress overshoot play a crucial role in the dynamics near
threshold. To do this, we study simpler models and hope that the
conclusions drawn from these models will help us understand the case
of real elastodynamics.

\subsection{Sharp Stress Pulses}

We first consider a simple model of the overshoots in which  sharp
stress pulse travels with the sound speed. We take the amplitude of
the overshoot to decay as a power
law of distance as it moves along the crack front. The continuum version 
of the kernel we study has the
form

\begin{equation}
{J}_{{\rm {sp}}}(z,t;\alpha ,\gamma )=\Theta (t-|z|)/z^{2}+\alpha
\delta (t-|z|)/|z|^{\gamma }  \label{nonmonotonicker}
\end{equation}
This kernel reduces to the previous case of the monotonic time delayed
interactions when $\alpha =0.$

In Figure[\ref{avpicovershoot}] a large avalanche that occurs on
triggering the most weakly pinned site is shown for both the monotonic
time delayed kernel and for the kernel $J_{{\rm sp}}(z,t;\alpha
=0.5,\gamma =1.5)$ starting from the same initial configuration of the
crack front and the same configuration of 
random toughnesses. We see that in the
presence of the overshoot many more sites are triggered than for the
monotonic kernel.  Our data show that even for small overshoots their
effects build and cause sufficiently large avalanches to run away. This
causes the crack front to de-pin and start moving at a threshold load
which is less than the one for the quasistatic stress transfer.

We find that for any value of $\gamma $ and any non-zero $\alpha $,
the threshold load, $ G_{c}(\alpha ,\gamma )$, is lower than
$G_{c}(\alpha =0,\gamma )=G_{c}^{qs}$. A plot of $\langle
{G_{c}^{qs}-G_{c}(\alpha ,\gamma )\rangle }$ as a function of $\alpha
^{2}$ is shown in Fig[\ref{thresh}] for various values of $\gamma $,
including $\gamma =\infty $, for which only the nearest neighbor of a
jumped site feels the overshoot.  The shift in threshold load from
$G_{c}$ was obtained by averaging over the {\it same} set of random
samples with and without the overshoot; this greatly reduces the error
bars. The value of $G_{c}$ can be accurately determined from the
quasistatic model where the code is much less numerically
intensive. For all the values of $ \gamma $ studied, $\gamma \geq
0.5$, we finds results consistent with
\begin{equation}
G_{c}^{qs}-G_{c}(\alpha ,\gamma )\sim \alpha ^{2}  \label{4alpha}
\end{equation}
for small $\alpha $, a form which will be derived below.  Thus the
overshoot appears to always be {\it relevant} at the quastatic
 depinning fixed
point.

One of the advantages of starting with monotonic models for which we
have quite a detailed understanding, is that the effects of
perturbations away from these can be analyzed using known scaling
properties of the monotonic models. We would like to carry this out
for weak stress pulses added to the time-delayed monotonic model;
i.e., to consider the behaviour of the crack with stress transfer given
by Eq.(\ref{nonmonotonicker}) for small $\alpha .$ In order to do this
we first obtain the response to a {\it single} stress overshoot. 

We
focus on a given space time point, at $(Z,T)$ which we denote ``$A$''
and an avalanche that started at ($0,0)$ in the time delayed model
without stress pulses. For simplicity we restrict consideration
initially to $ \gamma \rightarrow \infty $ so that only the nearest
neighbor of a jump site will feel a stress pulse which will be $\alpha
$ above the static stress.
In order for $A$ to be affected, one of the two neighboring sites of
$A$, say $Z-1$, must have jumped at time $T-1$ producing a pulse;
denote this space--time point $``P".$ Three conditions must be met for
$A$ to be affected by the stress pulse from $Z-1,$ i.e., for point $Z$
to jump an extra time.  First, $A$ must be within $\alpha $ of
jumping anyway for the stress pulse to have been able to trigger its
jump. The force increment needed for individual sites to jump are
uniformly distributed for sites near to jumping, so the probability of
this is simply of order $\alpha .$ Second, the increase in stress at
site $Z$ {\it  after} time $T$ must be less than $\alpha $, or else
the site would have jumped again regardless; denote this condition
``$L$''(for ``later stress'').  Third, the neighbor must jump at
$T-1,$ i.e., the jump $P$ must occur.

The scaling properties of avalanches imply that within the space-time
volume of a large avalanche, there are holes on all scales ( since
$l^{\zeta}l<<l^{2})$ and sub-avalanches of all sizes; see 
Fig[\ref{avmodqcomp}a].  Consider a sub-avalanche that occurs in a
region within a time $\tau <<T$ before $P$ and within distance of
order $ \tau $ from $P$ that is mostly contained within the
``backward sound cone'' of $P$, as shown in
Fig[\ref{alphaargument}]. If $P$ is not well enough ``isolated'' from
the bulk of this sub-avalanche, then we must consider a smaller
sub-avalanche until we find the largest $\tau $ such that there is a
sub-avalanche with size of order $\tau $ from which $P$ is
``isolated'' in space-time by order $\tau $ ( if this does not exist,
then it is highly unlikely that condition $L$ can be satisfied). By
isolated, we mean that there are few (or no) jumps on or outside of
the backwards sound cone of $P$ ---``$P^{\prime}$s cone''--- within a
time $\tau.$ This is illustrated in Fig[\ref{alphaargument}].  The
maximal such sub-avalanche which we denote ``$S$'' typically has of
order $\tau ^{1+\zeta }$ jumps of which a fraction $1/\tau $, i.e. $
\tau ^{\zeta }$, are on $P^{\prime}$s sound cone. Since these are
typically a distance $\gtrsim \tau $ from $Z$, they each cause the
stress at $P$ to increase by order $1/\tau ^{2}.$ Thus the isolated
space-time point $ P $ will typically have a stress increase of
$1/\tau ^{2-\zeta }$ ( between time $T-2$ and $T-1$) due to the
sub-avalanche $S$, the probability of a jump at $P$ given $S$ is of
this same order.

The condition that $P$ is isolated from $S$ is not a stringent
one. Any jump in $S$ outside of $ A^{\prime }s$ cone (\ which is
almost identical to $P^{\prime }$s cone) will cause a stress increase,
$\Delta _{L}$, at $Z$ after time $T.$ Since there will be of order
$\tau ^{1+\zeta }$ such jumps, each yielding a stress increase at $Z$
of order $1/\tau ^{2}$, $\Delta _{L}$ will be of order $ 1/\tau
^{1-\zeta }$. This stress increase will be larger than $\alpha $, and
hence violate condition $L$, unless $\tau >1/\alpha ^{1/(1-\zeta )}$
(note that obtaining this condition by either a smaller sub-avalanche
with anomalously few jumps outside of $A$'s cone or by $P$ being less
isolated from $S$ is very unlikely). The probability of both satisfying
the condition $L$ that the stress increase after $T$ be small {\it and}
having a jump at $P$, is thus controlled by the smallest $\tau ,$
\begin{equation}
\tau _{\alpha }\sim 1/\alpha ^{1/(1-\zeta )}  \label{183a}
\end{equation}
for which $L$ is satisfied with reasonable probability. The relevant
subavalanches $S$ are thus of size of order $\tau_{\alpha}$ and occur
in a space-time region of extent $\tau _{\alpha }\times \tau _{\alpha
}$ near $P$. Since $L$ is then quite probable
\begin{equation}
{\rm Prob}({\text {jump at $P$ {\it and} L$|$size
(S)$\sim\tau_{\alpha}$}})\sim 1/\tau _{\alpha }^{2-\zeta }\sim \alpha
^{1+1/(1-\zeta )}, \label{num83}
\end{equation}
which is simply proportional to the increase in stress at $P$ due to $S$.

What is the chance that there is such an appropriate sub-avalanche $S$
in the space-time region within $\tau _{\alpha }$ of $P$? If there is any
activity in this region, then it should include subavalanches on all
scales, but we also need this to be the {\it last} activity in this
region (or else $L$ will be violated). If $T$ is of order the duration
of the full avalanche, the last activity in this region could occur
anywhere within a time of order $T.$ Thus
\begin{equation}
{\rm Prob}({\text{last  activity being within $\tau_{\alpha}$ of $T$}})\sim
\tau_{\alpha}/T.  \label{184a}
\end{equation}
We obtain the probability of a pulse-triggered extra jump at $A$ by
 combining all the factors from Eq.(\ref {num83}),
 Eq.(\ref{184a}) and the probability $\alpha$ of site $Z$ being close
 to jumping again, yielding
\begin{eqnarray}
{\rm {Prob}}({\text{pulse triggering extra jump at $A|$ avalanche of size 
$\sim T$}}) &\sim &\frac{1}{\tau _{\alpha }^{2-\zeta }}\frac{\tau
_{\alpha }}{T}\alpha \nonumber \\
\sim  \frac{\alpha ^{2}}{T}
\end{eqnarray}
using the $\tau_{\alpha}$ given by Eq.(\ref{183a}). Since such an
event could occur over a range of times of order $T$ and the
number of sites in the original avalanche is of order $T$, the
total number of primary extra triggers caused by the pulses is of
order
\begin{equation}
N_{1}\sim \alpha ^{2}T
\end{equation}

The spatial density of the primary extra triggers is small so that
each of them will cause roughly independent secondary avalanches
under the dynamics without further stress pulses. The probability of
large avalanches falls off slowly with their size up to the
correlation length $\xi $, which we assume is bigger than $T$ so that
the original avalanche was not exponentially unlikely. This means that
the total number of jumps in these secondary avalanches will be
dominated by the largest one which has size
\begin{equation}
l_{{\rm {max}}}\sim (N_{1})^{1/\kappa }
\end{equation}
Then the total number of secondary jumps caused by the primary extra
triggers is 
\begin{equation}
M_{2}\sim (\alpha ^{2}T)^{(1+\zeta )/\zeta }
\end{equation}
using $\kappa =\zeta .$

Each of these secondary jumps has the potential of triggering more
jumps due to pulses. The number of such secondary triggers, $N_{2},$
will be much less than $N_{1}$ unless $M_{2}\sim M_{1}$;
the original number of jumps.  Thus, for fixed $T$, small enough
$\alpha $ will cause a small number of secondary jumps, fewer tertiary
ones, even fewer quaternary ones, etc.  But if
\begin{equation}
M_{2}\sim M_{1}\sim T^{1+\zeta }
\end{equation}
i.e. $T\sim \alpha ^{-2/(1-\zeta )}$, the process will run away. Thus we
expect that the stress pulses will become important when 
\begin{equation}
\xi \sim \xi _{\alpha }\sim \alpha ^{-2/(1-\zeta )}
\end{equation}
Using $1/\nu =1-\zeta $ , this corresponds to a reduction in the
critical load proportional to $\alpha^{2}$, i.e., of exactly the form
Eq.(\ref{4alpha}) that provided a good fit ot the numerical data.

Longer range pulses can be considered by a generalization of the above
argument.  It is found that the $\alpha ^{2}$ dependence is preserved
if and only if $ \int dzJ_{p}^{2}(z)<\infty $, with $J_{p}(z)$ the
peak pulse height (as a function of time) at distance $z.$ Our
results for $\gamma >1/2$ agree well with the predicted $ \alpha
^{2}$ dependence of the reduction in the critical load. The marginal
case $\gamma =1/2$ is similar numerically and we have not explored
smaller $ \gamma .$

For non-monotonic models, the no-passing rule discussed earlier and in
Appendix \ref{nopass},
 does not apply. Thus, at least for finite systems and for
some length of time, moving and stationary solutions
can co-exist. Indeed for any non-zero $\alpha $, we expect that for
loads in the range $G_{c}(\alpha ,\gamma )<G<G^{qs}_{c}$, both
stationary and moving solutions should co-exist and the selection
between the two will be determined by the past history. This effect is
seen in Fig[\ref{abruptstop}] in which the velocity of a crack front
of length 64, averaged over a cycle, is shown as a function of the
number of times it has passed through a sample of extent $W=16$ with
periodic boundary conditions in the direction of motion. In monotonic
models, convexity implies that at long times we would measure the same
velocity in every pass as there is a unique steady
state{\cite{nopassing}}. But with stress pulses, we see that the
velocity changes with the cycle, there is no unique moving solution,
and after a number of cycles the crack front can suddenly come to a
complete halt, as in Fig[\ref{abruptstop}], thereby directly
illustrating the co-existence of moving and stationary solutions. In finite 
systems with periodic boundary conditions in both directions as we have 
used,
 whether all moving states eventually stop or whether they can survive 
indefinitely
is likely to depend both on the sample and on the load. The question of what 
happens for 
infinite systems, we return to in the last section.

\subsection{Scalar Elastic Approximation}

We finally consider kernels of the form
\begin{equation}
{J}_{{\rm {se}}}(z,t;\tau )=\frac{(t+\tau_{0})\Theta (t-|z|)}{\pi
z^{2}((t+\tau_{0} )^{2}-z^{2})^{1/2}}  \label{Jtau}
\end{equation}
which is appropriate for the scalar approximation to
elasticity. Unlike the sharp pulse models the stress overshoot has a
long tail in time. We have cut off the singularity at the sound
arrival time by a time $\tau_{0} $ which crudely represents the
microscopic response time. For $\tau_{0} \rightarrow \infty $, this
model becomes the monotonic model with the time delayed kernel.

We measure the threshold load as a function of $\tau_{0} ,$ for
large $ \tau_{0} ,$ and find the reduction of the critical load
$\langle G^{qs}_{c}-G_{c}(\tau_{0} )\rangle $ is consistent with
$\tau_{0} ^{-3/2}$as shown on a log-log plot in Fig[\ref
{thforcewitho}], we do not, however,
 have an analytical argument for the
exponent $3/2.$ Not surprisingly, we again find that the
non-monotonicity of the kernel is relevant but with a larger
eigenvalue that would be expected from the peak pulse heights,
presumably because of the long time for which the overshoots are
substantial.

\subsection{Velocity and Hysteresis}
For both the sharp pulse and the scalar elastic models, the stress
overshoots lead to velocity versus load curves that are both very noisy
and hysteretic. For the scalar elastic case, results are shown for
various $\tau_{0}$ in Fig[\ref{hysteresis}].  As the load is
increased, the velocity appears to jump to a non-zero value which is a
function of the overshoot's strength and then jump back to zero again
on decreasing the load only at smaller load. Thus the stress
overshoots {\it appear} to lead to a first order like transition, where the
crack front jumps directly to a finite velocity from the pinned
phase. Finite sized systems show hysteretic behaviour as shown in
Fig[\ref{hysteresis}]. Several cautionary remarks are, however, in
order. First, getting statistics on the sizes of hysteresis loops for
these non-monotonic models is numerically intensive. As we saw in
Fig[{\ref{abruptstop}], due to the non-monotonic nature of the models
the crack front can come to a complete halt after passing through the
system several times. The fluctuations in the mean velocity per cycle
of the crack front increases with the size of the stress overshoot but
for a given magnitude of the overshoot, decreases as the system size
increases. The abrupt stopping of the crack front naturally leads to a
large scatter in the size of the hysteresis loops, particularly for
smaller systems. Thus, even though the computation time per step
increases with system size as $L^{2}{\rm ln}L$, the scatter in data of
the hysteresis loop sizes for small system sizes makes it difficult to
study the system size dependence of the hysteresis loops. Second, it is
not possible to ascertain whether the crack front will eventually stop
or not unless the load is above the $G_{{\text{min}}}^{moving}$ of the
quasistatic model; i.e., the largest load at which a static solution
exists. Third, it should be noted that the minimum velocities both on
increasing and on decreasing the load in Fig[\ref{hysteresis}] are not
all that much bigger than the quasistatic $v_{{\text{min}}}$, even for
$\tau_{0}=1$ for which $G_{c}$ has decreased by almost a factor of
two. Thus, overall, it is not clear at this point which of the effects
that are apparent in the numerics for these non-monotonic models are
finite-size effects and which are indicative of the behaviour of much
larger systems.  We return to this issue at the end of the paper.

\section{Discussion}

\label{discussion3} 

In this last section we compare our results on the dynamics of planar crack 
fronts with other work and discuss various open questions.

\subsection{Quasistatic Limit}

In the absence of sound waves, long range
elasticity leads to a  non-local but monotonic stress transfer kernel
in the equation of motion of the front. The transition
from the pinned to the moving phase is then second order and there is a unique 
moving solution above
threshold. There are two independent critical exponents in this case and the
numerical results we obtain using a discrete model are in good agreement with 
those from the
$ \varepsilon $- expansion. Note, however that it was necessary to include the 
effects of corrections to scaling both to get reliable estimates of the 
exponents and to verify the scaling laws. The dynamic exponent $z$ and the 
velocity exponent $\beta$ are both
found to be less than one as predicted by the $ \varepsilon $- expansion. The 
roughness exponent $\zeta$ is very close to the lower bound of $\frac{1}{3}$ 
which may well be exact, although we have no solid argument for this.

There have recently been two other numerical studies on quasistatic crack 
models with the appropriate long range interactions.  The first, by Schmittbuhl 
{\it et.al}\cite{schnum}, obtains the
same value of the roughness exponent $\zeta$ as we do to within error
bars. However they obtain a dynamic exponent which is greater than
one. It is not clear from their paper as to how this result was obtained and  
at this point the discrepancy is not understood. The second
paper, by Thomas and Paczuski \cite{pac}, obtains $\zeta \approx \frac{1}{2}$. 
Their system sizes are large and it is not at all clear why the results should 
be so different. One possibility is the dynamical ``updating rules''. 
Thomas and 
Paczuski's are different and
more unphysical than ours. Rather than increasing the force
adiabatically to depin the most weakly pinned point on the crack
front, they depin the point on the crack front that is
farthest behind the rest. Whether this or some other difference is the cause of 
the differences we leave as an unresolved question.

In addition to these numerical studies, there is a very recent experiment 
\cite{schmexp} in which two halves of a block of plexiglass which have been 
roughened and then pressure welded together are broken apart.  The crack is 
thus confined to the plane of the original weld.  The crack front roughness is 
measured while it is advancing at a very slow mean speed. With a rather 
limited range of data, the authors obtain $\zeta = 0.55 \pm 
0.05$.  This would appear to be inconsistent with our quasistatic result, but a 
systematic curvature appears to be observable in the data and it may be 
popssible to fit it reasonably well with $\zeta = \frac{1}{3}$ 
and a $(1/l)^{-1/3}$ 
correction to scaling (a form with the same number of parameters as an unknown 
$\zeta$).  Another possibility, however, is that the experiments are not really 
in the quasistatic regime.

These experiments force us to address an issue which we have so far avoided: 
what determines whether (or in what regime) a crack will behave 
quasistatically?
The basic criterion is {\it not} directly related to the average velocity of 
the crack.
Rather, it is the speed of propagation of disturbances {\it along} the crack 
front - in particular this speed relative to the Rayleigh wave speed, $c$ - 
that is the essential determining feature.  But what determines the speed of 
propagation of disturbances is rather subtle.
In general, if the materials that make up the heterogeneous medium are 
themselves resonably close to ideal elastic solids, then there is no natural 
parameter which would make the speed of propagation of disturbances along the 
crack front much slower than $c$, even if the heterogeneities are weak. But if 
there is substantial plasticity, creep, or other dissipative effects on the 
scale of the heterogeneities, then even on these mesoscopic scales the 
equation of motion of the crack front is {\it not} given simply by the 
``propagator'' 
Eq.(\ref{eq18}) with $B$ its ideal value of order $1/c$. If the 
system is {\it velocity toughening} due to these small scale effects, i.e., the 
effective fracture toughness on the scale of the heterogeneities increases with 
velocity - or equivalently that the velocity increases more slowly with load 
than in an ideal solid - then the linearized dynamics is given by 
Eq.(\ref{eq18}) with a larger value of $B$.  In the limit of very large $B$, 
the large scale behavior of the crack front will be well approximated by 
the quasistatic model except very close to the onset of crack propagation
 where the 
cumulative effects of the stress pulses caused by a large avalanche will still 
cause it to run away.  Since the exponent $z$ is not much smaller than one, 
however, the crossover to fully dynamical behavior will occur only very close 
to the critical load.

Other non-linear effects - in particular those associated with the local 
depinning of a section of the crack front caused by the advance of other 
sections of the crack - will also affect how quasistatic the behavior of the 
crack front will be.  A careful study of several experimental systems - 
including investigating  whether or not the motion appears in bursts of 
activity when the crack is moving at a slow average speed - would appear to be 
needed to help resolve this and other related issues.

\subsection{Elastodynamic Effects}

Close enough to the critical load the quasistatic approximation always
breaks down. Since in this approximation the dynamic exponent $z<1$,
the effective propagation velocity of the disturbance associated with
a quasistatic avalanche of size $\xi$ diverges as $\xi^{1-z}$, thereby
becoming of order the sound speed sufficiently close to the critical
load no matter how small the ``bare'' velocity of small scale
disturbances.  Thus elastodynamic effects {\it must} alter the
asymptotic critical behavior.

In order to understand the effects of sound-travel time delays, we
first considered a simplified causal model in which the dynamic stress
transfer is still monotonic.  In this model, the static exponents
(i.e. $\zeta$ and $\nu$ ) were found to be the same as in the
quasistatic approximation because of the monotonic character of the
stress transfer. But the dynamic exponent, $z$, became equal to one,
the minimum value consistent with causality. The scaling identities
then imply that $\beta =1$ which is in good agreement with our
numerical results on this sound-travel-time-delayed model.

But the actual dynamic stress transfer along a crack is more
complicated.  Indeed, proper inclusion of the dynamics of the medium
necessarily leads to a non-monotonic kernel in the equation of motion
of a crack front. In particular, the stress that arises at a point on
the crack front due to another section of the crack moving forward,
generically rises to a peak before decaying to its long time
quasistatic value.

We have examined the effects of these stress overshoots and find that
they are {it always relevant} at the depinning transition.
Specifically, for sufficiently large avalanches the effects of the
overshoots build up enough to make such avalanches run away. This
causes the crack front to move - at least by some amount - at a load
which is {\it lower} than the quasistatic critical load, i.e., for
loads at which there are still stable static configurations.  (This
can occur because, in the presence of stress overshoots, the
``no-passing'' rule which prevented moving and stationary solutions
coexisting in monotonic models is no longer valid .)

If the stress overshoots are weak, their effects will not be important
until very close to the quasistatic critical load and there will be a
wide regime of validity of the quasistatic results.  They will
eventually break down only when the correlation length exceeds a
crossover length scale which has an inverse power law dependence on
the magnitude of the stress overshoots.

What happens when a large avalanche runs away?  We have explored this
by numerical studies using simplifed stress transfer kernels which
include both sound-travel-time delays and stress overshoots. In order
to investigate hysteretic effects, we have used finite systems with
periodic boundary conditions in the direction of motion with the
extent in this direction proportional to the cube root of the length
of the crack front - i.e, the scaling of the crack distortions at the
quasistatic critical load. In the absence of stress overshoots, moving
configurations converge to a unique periodic state. But, due to the
non-monotonicity, this need not be the case once stress overshoots are
included.

Nevertheless, for a range of loads between the point $G_{run}$ at
which an avalanche runs away and the quasistatic critical load
$G_c^{qs}$, we find that the crack front usually converges to a state
which is periodic in time with a period which is some multiple of the
time to pass through the system once. If the load is then decreased to
below $G_{run}$, the resulting moving state coexists with static
configurations which are {\it stable} to avalanche runaway under small
increases in the load. At still lower loads the behavior tends to
becomes chaotic, with, in at least some samples, the crack eventually
coming to rest only after passing through the sample many times as
illustrated in Fig[{\ref{abruptstop}].

The data we have collected thus suggests hysteretic behavior with
coexisting moving and stable stationary regimes coexisting in some
range of loads from $G_{run}$ down to some lower critical load,
$G_{stop}$. However, our numerical results indicate that the widths of
the hysteresis loops are quite a bit smaller than the difference
between the critical load $G_{run}$ and the the quasistatic critical
load. We would of course like to know whether the hysteretic behavior
persists in an infinite system. Unfortunately, the numerics are rather
slow in the presence of sound-travel-time delays. Thus, obtaining
statistics for the sizes of hysteresis loops is numerically intensive
and our results are far from conclusive.

\subsection{Possible Scenarios near Threshold}

In this last section we consider various possible scenarios for the behavior of 
large systems in the presence of stress overshoots.

The simplest scenario is suggested by our data: As the loading is
increased slowly, the stationary crack jumps to a non-zero velocity at
load $G_{move}$, but when the load is decreased, the crack does not
stop until a lower critical load $G_{stop}$.  At $G_{stop}$ the
velocity could either drop to zero discontinuously, presumably the
result of an instability of the moving phase, or continuously (as
occurs in a hysteretic underdamped Josephson junction). If the load is
changed suddenly, this could cause a jump from one phase to the other,
perhaps even in the range {\it above} $G_{move}$ but below $G_c^{qs}$
in which static configurations still exist . Note that the obvious
guess, suggested by our numerics, is that $G_{move}=G_{run}$, the
point at which avalanches run away and become much bigger than their
size in the absence of the stress overshoots.  However, it is not
obvious that this has to be the case: One could imagine a scenario in
which the runaway avalanches eventually stop but only after causing
the crack front to become much rougher than it would be under the
quasistatic avalanches.  A true moving phase might then only exist
above a higher load $G_{move}$.  This appears rather unlikely and
seems difficult to reconcile with our numerical simulations even
though the latter may have been biased by our choice of scaling of the
length and the extent of the finite systems.

A second scenario is that, in the limit of large system size, the
hysteresis loops we found numerically disappear and the transition
becomes ``first order '' with a discontinuous velocity versus load but
no hysteresis if one waits a long enough time for the crack to settle
down.

Finally, and perhaps most interesting, is the possibility that the
onset of crack motion could still be critical with the velocity rising
continuously at a critical load and some kind of diverging correlation
lengths as the critical load is approached adiabatically from above
and from below. This would represent a new universality class of
depinning transitions. A variant of this, with the velocity
discontinuous but the transition still critical in the sense of
diverging correlation lengths, is also conceivable.

Which of the above scenarios obtains may well depend on aspects of the
physics that we have left out of our numerical studies and theoretical
analysis. For example, multiple scattering of elastic waves from the
crack front will cause different regions of the crack front to see
stress pulses that depend on the shape of the crack front in their
vicinity and that of the segment which has moved.  However, we
conjecture that the general role of stress overshoots should not be
qualitatively changed by multiple scattering since the long wavelength
sound waves will not be strongly affected by the crack roughness
unless $\zeta=1$.

A potentially more important effect is a consequence of vectorial
elastodynamics.  In particular, for a tensile crack the behavior may
be complicated by the fact that the initial stress pulse caused by a
section of the crack jumping forwards is {\it negative} with the
stress only becoming positive when the Rayleigh waves arrive. If one
hypothesizes a hysteretic velocity-load curve, then in the hysteretic
region such a stress transfer kernel can support the coexistence of
moving and stationary zones of the crack front with the boundary
between the zones moving at a velocity $s$ that corresponds to the
zero of the kernel $P(k,\omega )$ ( for a Poisson's ratio of
$\frac{1}{4}$, $s\approx 0.94 c$). In the moving phase, an anomalously
tough region could thus cause a stopping ``shock'' to move along the
crack front.  This might result in a complicated - and very rough -
moving state involving large scale stopping and starting.  What roles
such shocks might play in the onset of macroscopic crack motion we
leave as an interesting avenue for future study.

Finally, we have totally ignored all effects of non-planar crack
deformations.  These almost certainly play a major role in many
experimental situations and may well be important whenever the crack
is not confined to a pre-weakened plane.

{\it Acknowledgements}

We would like to thank O. Narayan, J.~R.~Rice and D~.~Erta\c s for
useful discussions.  This work has been supported in part by the NSF
via DMR-9106237, 9630064 and Harvard University's MRSEC.

\appendix
\section{}
\label{nopass} 
In this appendix we discuss the {\it no-passing rule}\cite
{nopassing} in the context of the quasistatic model and show that the
monotonic model with the time delayed interactions and the quasistatic
models should have the same static exponents.

The elastic force in the quasistatic model can be derived from an elastic
potential defined as 
\begin{equation}
V(\{f\})=\int dk|k||f(k)|^{2}
\end{equation}
where $k$ is the wave vector corresponding to the $z$ co-ordinate and 
$f(k)$
is the Fourier transform of $f(z)$. This potential is convex in $f.$
Following Ref.\cite{nopassing} consider two configurations of the crack
front, $f^{G}(z,t=0)$ and $f^{L}(z,t=0)$ with $f^{G}(z,t=0)\geq f^{L}(z,t=0)$
\ $\forall \;z.$ The no passing rule states that this inequality holds for
all times. This can be seen by noticing that if the two configurations were
to pass, $f^{L}$ would have to first approach $f^{G}$ at some point $z.$ 
At
this point, the random fracture toughness and the external driving force
would be identical for both the conformations of the crack front. However,
the elastic forces at $z$ on $f^{L}$ would be less than or equal to that on 
$
f^{G}$at $z$, due to the convexity of the potential. This prevents the
passing of $f^{G}$ bby $f^{L}$ and hence the inequality $f^{G}(z,t)\geq 
f^{L}(z,t)$ is obeyed at all times. The no
passing rule also implies a unique moving solution for the crack front in
the quasistatic model. \cite{nopassing}.

Now consider the monotonic model with the time delayed interaction. We 
see
that 
\begin{equation}
J_{{\rm {td}}}(z,t)\leq J_{{\rm {qs}}}(z,t)\;\ \forall \;(z,t)
\end{equation}
where $J_{{\rm {td}}}(z,t)$ and $J_{{\rm {qs}}}(z,t)$ are the kernels
describing the elastic interactions in the two models. This inequality
holds when we define the kernel with the sound-travel-time delayed 
interactions at $z=0$ as in Eq.(\ref{delayedmodq}). Thus we see that if
we consider two crack fronts, $f_{{\rm {qs}}}(z,t)$ obeying the quasistatic
equation of motion and $f_{{\rm {td}}}(z,t)$ obeying the monotonic time
delayed equation of motion, with $f_{{\rm {qs}}}(z,0)=$ $f_{{\rm {td}}
}(z,0)\;\forall \;z$, then following the previous argument we see that $f_{
{\rm {qs}}}(z,t)\geq $ $f_{{\rm {td}}}(z,t)\;\forall \;(z,t).$ Also since,
as $t\rightarrow \infty ,$ $J_{{\rm {td}}}(z,t)\rightarrow J_{{\rm {qs}}
}(z)\;\forall \;(z),$ we see that, if the load is below threshold, at
the end of the avalanche  $f_{{\rm {qs}}}(z)=$ $f_{{\rm {td}}}(z).$ Thus if
we start with the same initial configuration, the final positions of the
crack front at the end of each avalanche are identical. Thus the static
properties are the same for both the models.

By defining the kernels as in Eq.(\ref{principal}), we make sure that both
our scalar model as $\tau_{0} \rightarrow \infty $ and the sharp pulse 
model
with $\alpha =0$ behave like the monotonic model with the time delayed
interaction.

\section{}
\label{kernelsnum}
In this appendix we give the explicit forms of the interaction
kernels of the various models that we studied numerically.

For the quasistatic approximation, the discretized version of the 
interaction
kernel, Eq.(\ref{jqs}), 
with periodic boundary conditions in the direction along the crack front
 is given, as in Eq.(\ref{jqsdis}), by
\begin{equation}
\tilde{J}_{qs}(z)=\frac{1}{||z||^{2}}
\end{equation}
for $z\neq 0$ and 
\begin{equation}
\tilde{J}_{qs}(z=0)= - \sum\limits_{z\neq 0}\tilde{J}_{qs}(z),
\end{equation}
where 
\begin{equation}
||z||\equiv min(|z|,|L-|z||).
\end{equation} 

In the case of the monotonic model with the time delayed interactions, 
the discretized version of the interaction kernel Eq.(\ref{jtd}) is given by
\begin{equation}
\tilde{J}_{td}(z,t)=\frac{1}{||z||^{2}}\Theta(t-||z||)
\end{equation}
for $z\neq 0$. As noted in the text and in the previous appendix, there is 
an ambiguity in defining the kernel at $z=0$, and in order to preserve the 
monotonicity properties of the kernel we define
\begin{equation}
\tilde{J}_{td}(z=0,t)= - \sum\limits_{z\neq 0}\tilde{J}_{qs}(z)
\end{equation}

The other models we consider in this paper have stress overshoots which 
decay
as the stress pulse moves along the crack front. To include these effects with 
periodic
boundary conditions, we design our 
kernels such
that the stress overshoots disappear smoothly after running through 
the 
system once, after which time the kernel equals the quasistatic kernel, 
$\tilde{J}_{qs}$ at all
points in space. Thus, for the sharp pulse model defined by the kernel Eq.
(\ref{nonmonotonicker}), we choose the discretized kernel to be
\begin{equation}
\tilde{J}_{sp}(z,t\leq 
L/2)=\frac{1}{||z||^{2}}\Theta(t-||z||)+\alpha\frac{\delta(t-||z||)
}{||z||^{\gamma}}e^{-\frac{1}{L/2-t}}
\end{equation}
and
\begin{equation}
\tilde{J}_{sp}(z,t\geq L/2)=\frac{1}{||z||^{2}}
\end{equation}
for $z\neq 0$. There is again an ambiguity as to how one chooses the kernel at 
$z=0$ and we have defined it to be
\begin{equation}
\tilde{J}_{sp}(z=0,t)=-max[\sum\limits_{z\neq 0}\tilde{J}_{sp}(z,t),
\tilde{J}_{qs}(z=0)].
\end{equation}

Finally, in the case of the scalar model, the discretized form of the kernel, 
Eq.(\ref{44scalarkernel}), is chosen once again such that the overshoots 
vanish
after they have run through the system once. Thus,
\begin{equation}
\tilde{J}_{se}(z,t\leq L/2)=\frac{t+\tau_{0}}{||z||^{2}}\frac{1}{[
(t+\tau_{0})^{2}-||z||^{2}e^{-\frac{1}{L/2-t}}]^{1/2}}
\end{equation}
and
\begin{equation}
\tilde{J}_{se}(z,t\geq L/2)=\frac{1}{||z||^{2}}
\end{equation}
for $z\neq 0$.
As for the sharp pulse model, we define
\begin{equation}
\tilde{J}_{se}(z=0,t)=-max[\sum\limits_{z\neq 0}\tilde{J}_{se}(z,t),
\tilde{J}_{qs}(z=0)].
\end{equation}

\begin{figure}
\begin{center}
 \epsfxsize=5in
\epsffile{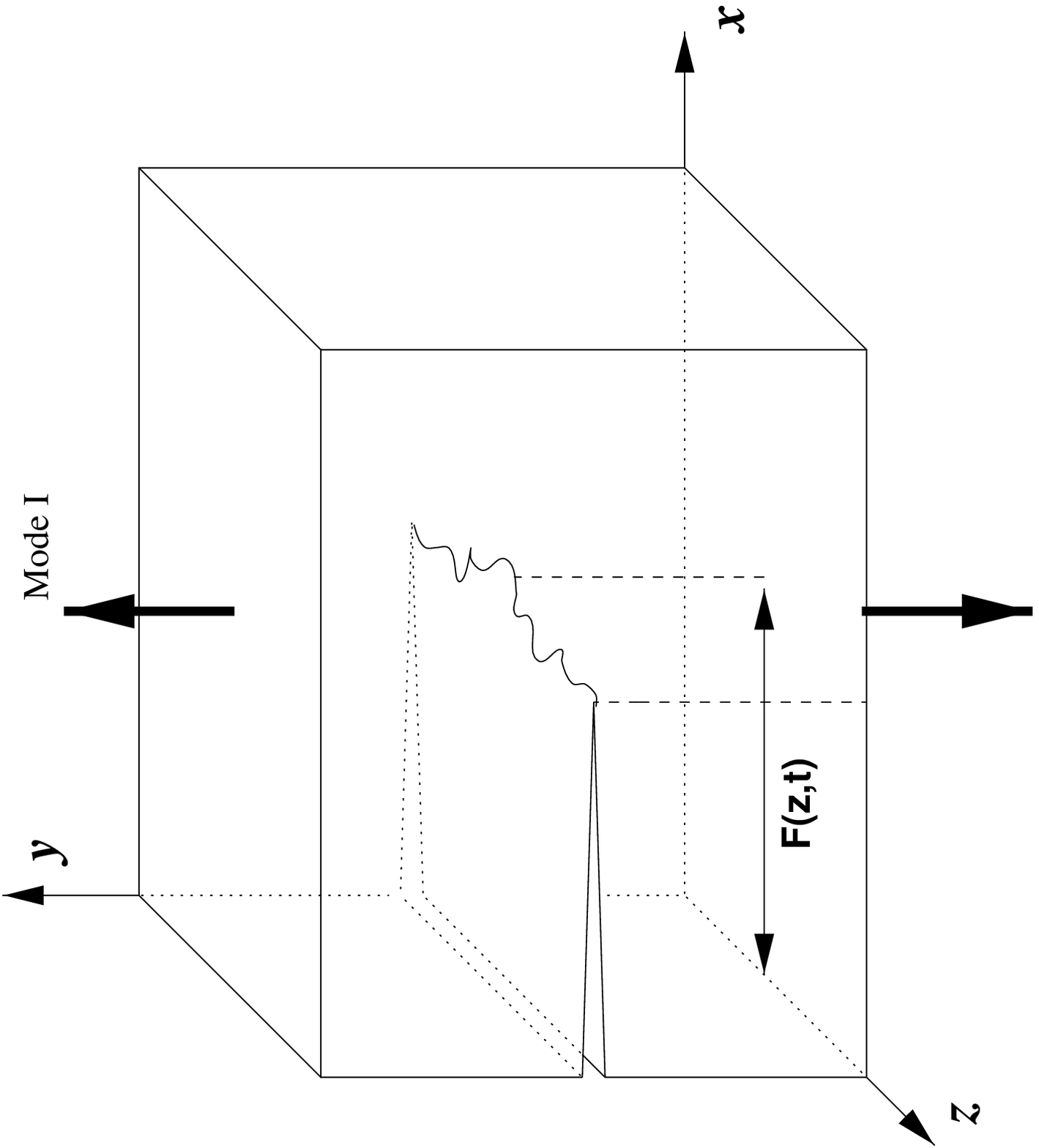}
\medskip
\caption{Schematic of a planar crack propagating through a 
heterogeneous medium. The crack
front $x=F(z,t)$ and the free crack surfaces, which are flat, are shown. The
applied mode I (tensile) loading is also indicated.}
\label{4crackgeom}
\end{center}\end{figure}

\begin{figure}
\begin{center}
 \epsfxsize=5in
\epsffile{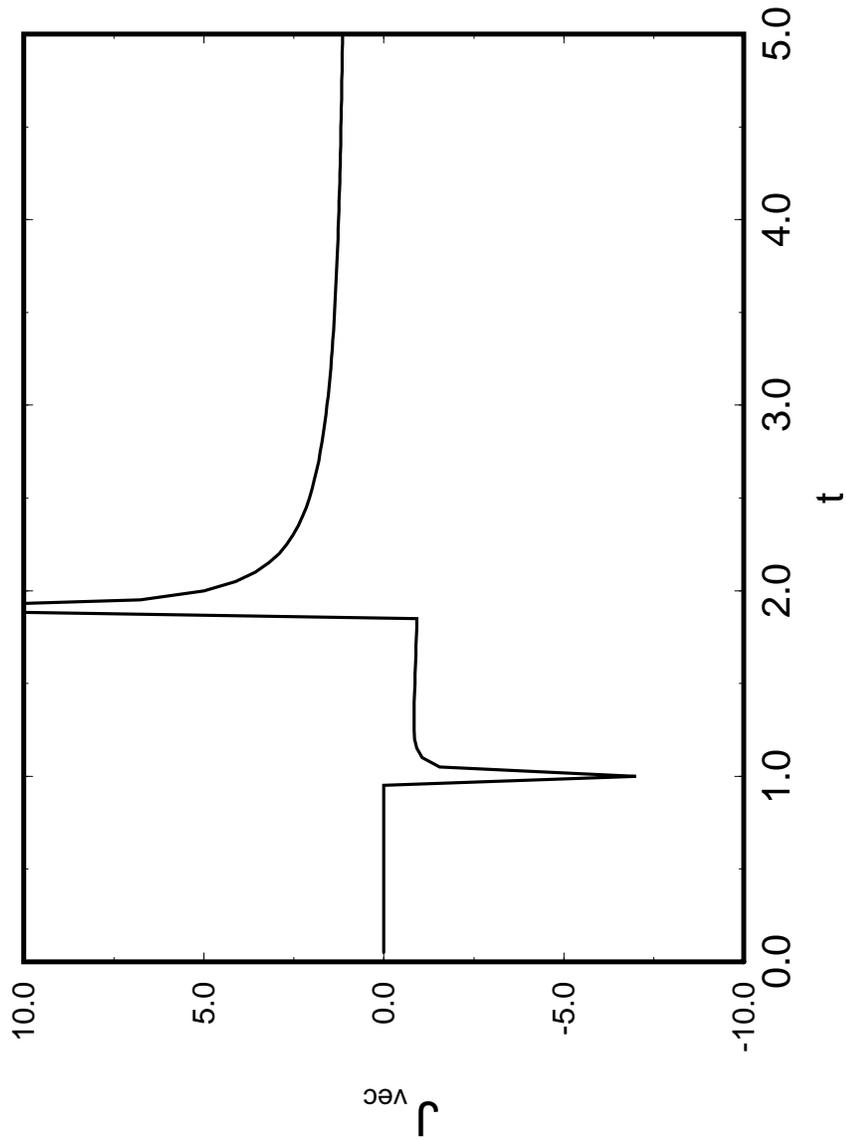}
\medskip
\caption{Stress transfer kernel for a tensile crack
 is shown as a function of $t$ at fixed distance $z=1$. The longitudinal
sound velocity is set equal to one and the Poisson ratio chosen to be 
$\nu =0.25$. The stress pulse is initially negative and then
changes sign. The divergences at times corresponding to the longitudinal 
and
Rayleigh wave arrival times have been cut off.}
\label{fig:jvec}
\end{center}\end{figure}

\begin{figure}\begin{center}
 \epsfxsize=5in
\epsffile{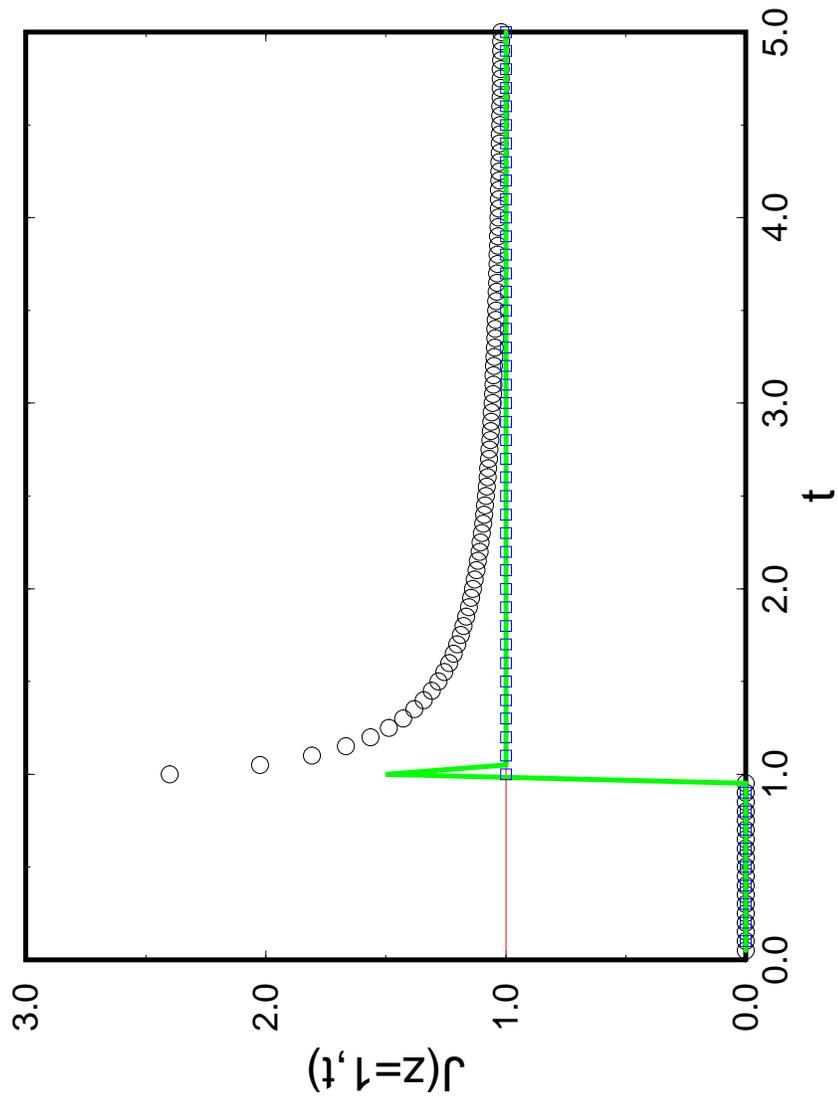}
\medskip
\caption{Stress transfer kernels, $J$, for various
models are shown as a function of time at a fixed distance
$z=1$. The thin line is the kernel for the
quasistatic model and the squares that for the sound-travel-time delayed
monotonic model. The open circles represent the scalar elastic model with 
$\tau_{0}
=0.01$ while
the thick line represents the sharp pulse model with $\alpha=0.5$ and $
\gamma=1.5$.  For all except the quasistatic model, 
the stress transfer is zero for $t<z$. }
\label{pulseplot}
\end{center}\end{figure}

\begin{figure}\begin{center}
 \epsfxsize=4in
\epsffile{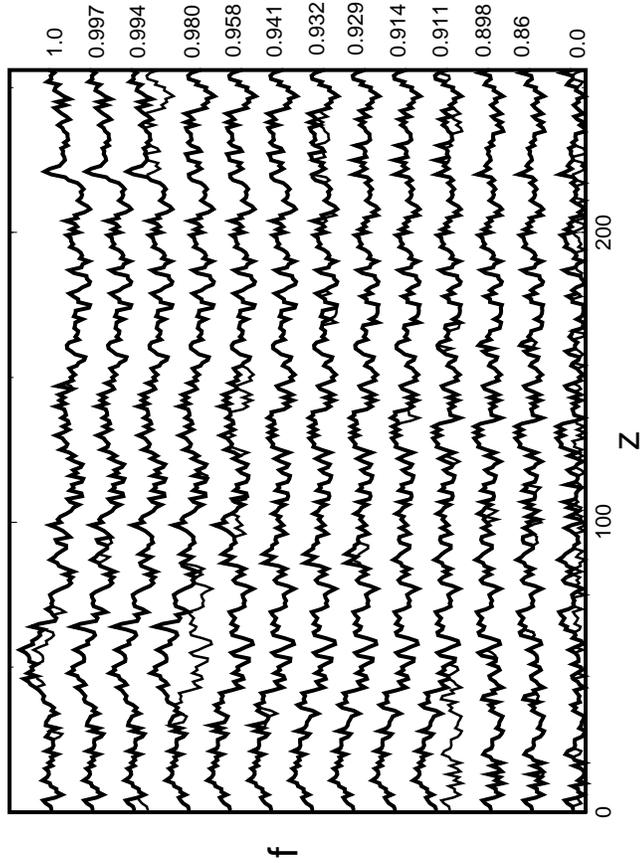}
\medskip
\caption{Series of avalanches in the quasistatic model in 
a system of size 256, as the
load is gradually increased by just enough at each step
 to trigger the most weakly pinned site. The
system is then allowed to evolve until motion stops
 before the load is increased again. The
configuration of the crack front is shown by a thin line at the begining and
by a thick line at the end of each avalanche to demarcate the sites which
have moved. The position of the crack front in the figure
is displaced vertically 
by a constant factor after each avalanche to
differentiate between the individual avalanches. The initial almost straight 
configuration of the crack front at zero load is also
shown. The avalanches shown occur
at fractional loads (indicated on the right) in the range from $0.86$ of
the threshold load to the threshold load at which point the crack front 
starts
moving.}
\label{avmodq}
\end{center}\end{figure}

\begin{figure}
\begin{center}
 \epsfxsize=5in 
\epsffile{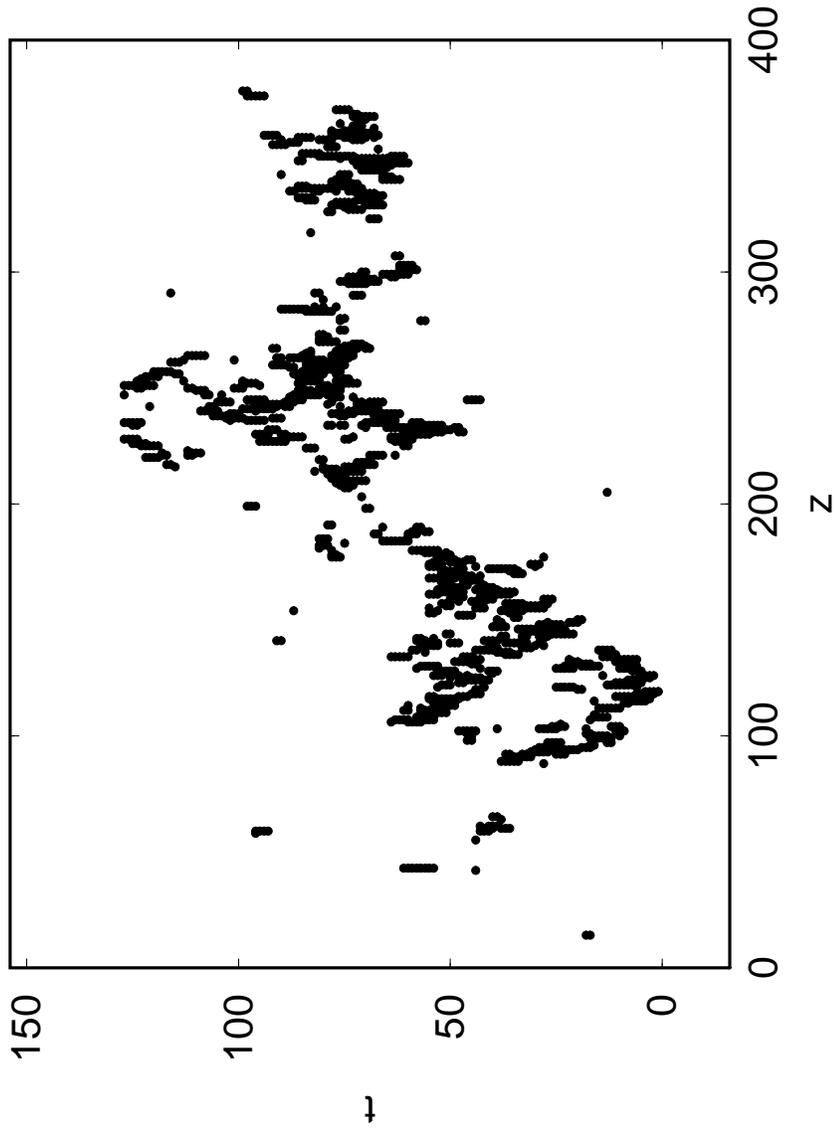}
\medskip
\caption{Space-time plot of a large avalanche in the quasistatic model
in a system of size 512, with
 the points on the interface which moved ahead at each
instant of time indicated.}
\label{avpic}
\end{center}
\end{figure}

\begin{figure}\begin{center}
\epsfxsize=5in
\epsffile{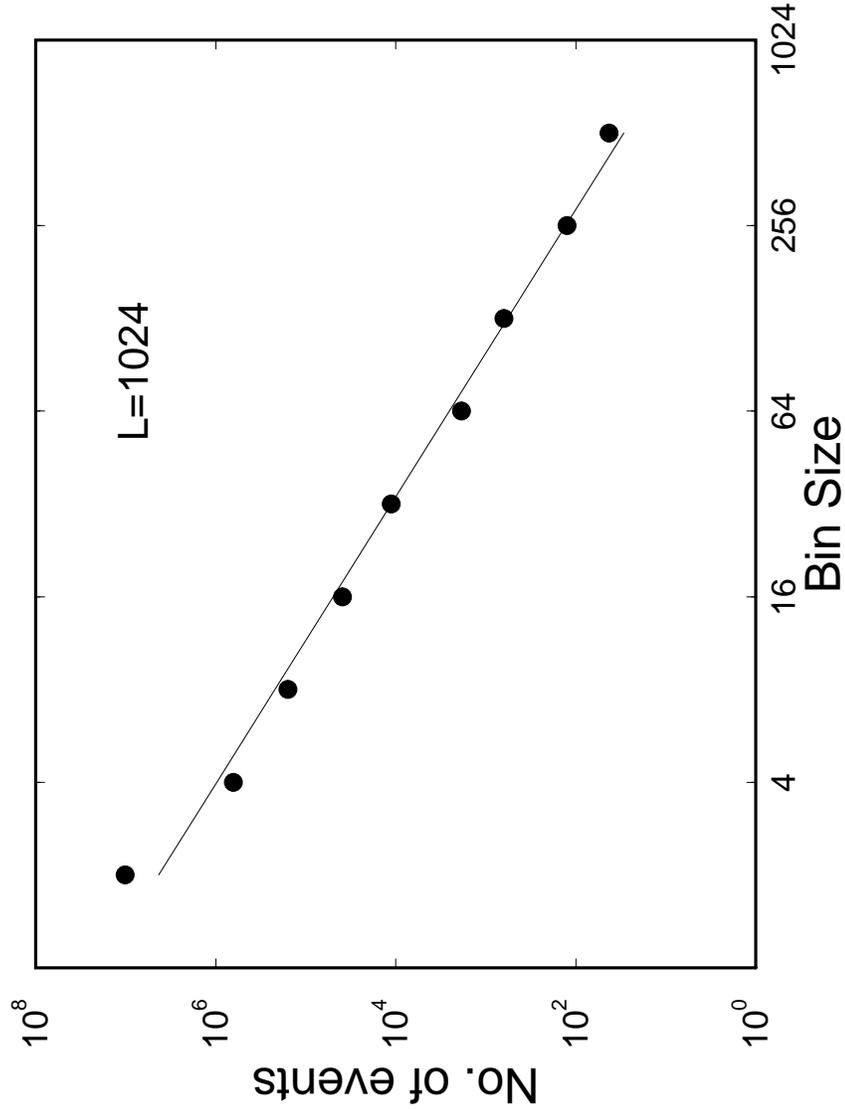}
\medskip
   \caption
  {As the load is increased from zero towards the critical load, the 
avalanches that occur in the quasistatic model, are binned according to
their size, which is defined as the number of distinct 
sites that move during an
avalanche. The bin is defined by $({\rm bin\ size})/\sqrt{2}<{\rm avalanche\ 
size}\leq \sqrt{2}\ ({\rm{bin\ size}})$. 
The number of avalanches in each bin are plotted versus the
bin size for systems of 
 size 1024. The slope of the log-log plot is 2.14$\pm 0.3$.} 
  \label{fig:binsize}
  \end{center}\end{figure}

\begin{figure}\begin{center}
\epsfxsize=5in
\epsffile{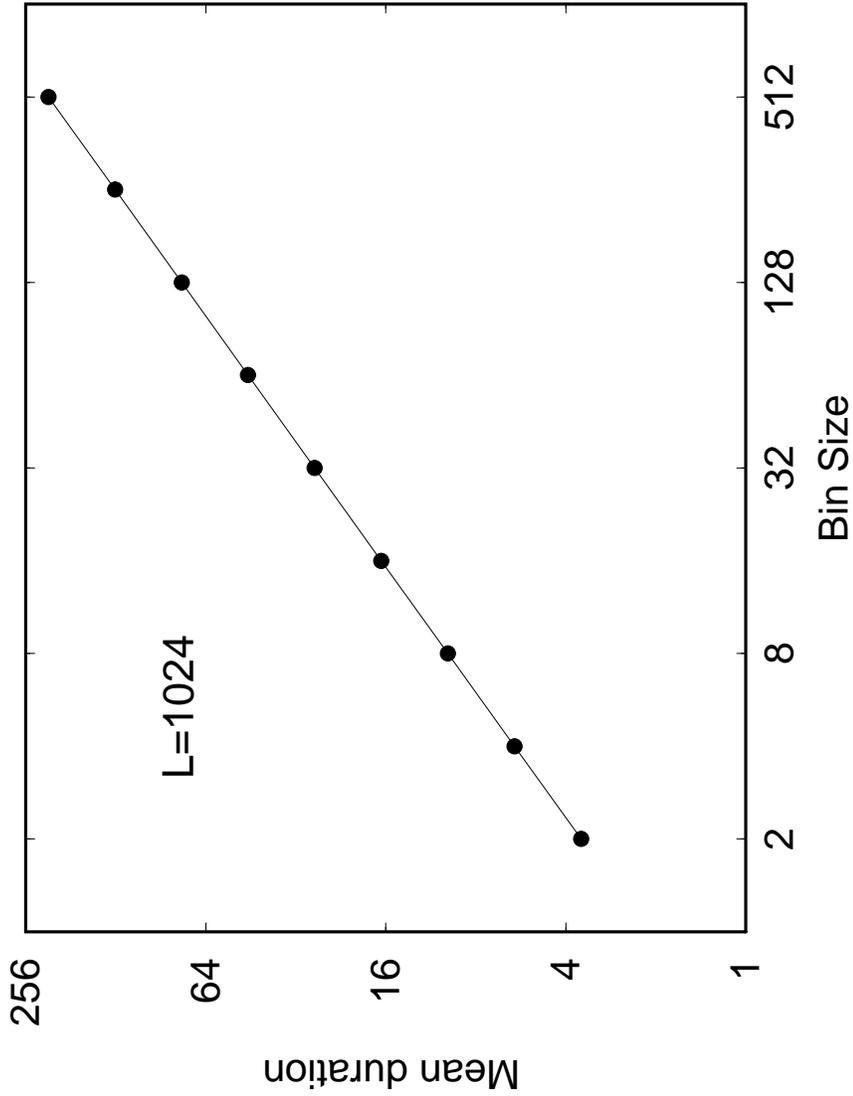}
\medskip
   \caption
  {Log-log plot of  the mean duration of an avalanche versus the bin size
is shown for the quasistatic model with system size 1024. The avalanches 
are binned according to
their size which is determined by the number of sites that moved during 
that
avalanche. The nth bin is defined by  $({\rm bin\ size})/\sqrt{2}<
{\rm avalanche\ 
size}\leq \sqrt{2}\ ({\rm{bin\ size}})$. 
The slope of the graph yields a dynamic critical exponent of $z=0.74\pm 
0.03$.}   
  \label{zexp}
  \end{center}\end{figure}

\begin{figure}\begin{center}
\epsfxsize=5in
\epsffile{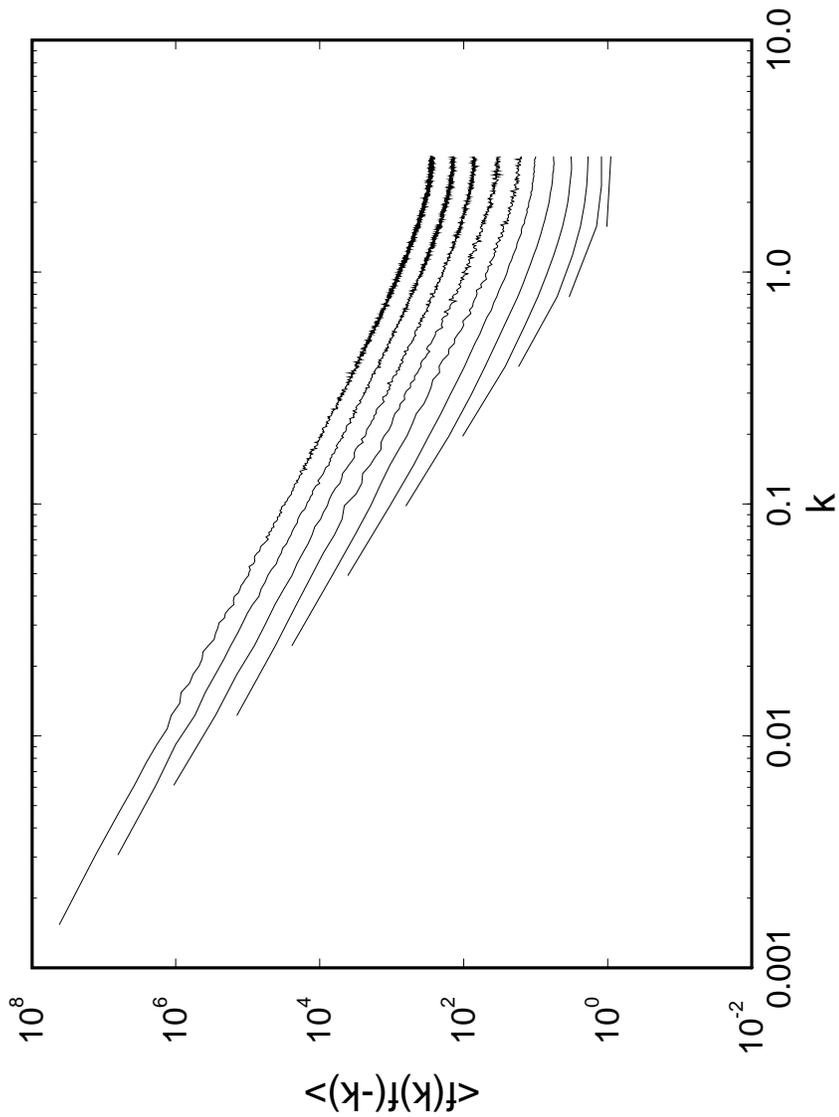}
\medskip
   \caption
  {Log-log plot of the power spectrum versus the wave vector is shown
for systems at the quasistatic 
critical load, ranging in size from 4 to 4096. They have been shifted along
the vertical axis for clarity.}   
  \label{zeta}
 \end{center}\end{figure}

\begin{figure}\begin{center}
\epsfxsize=5in
\epsffile{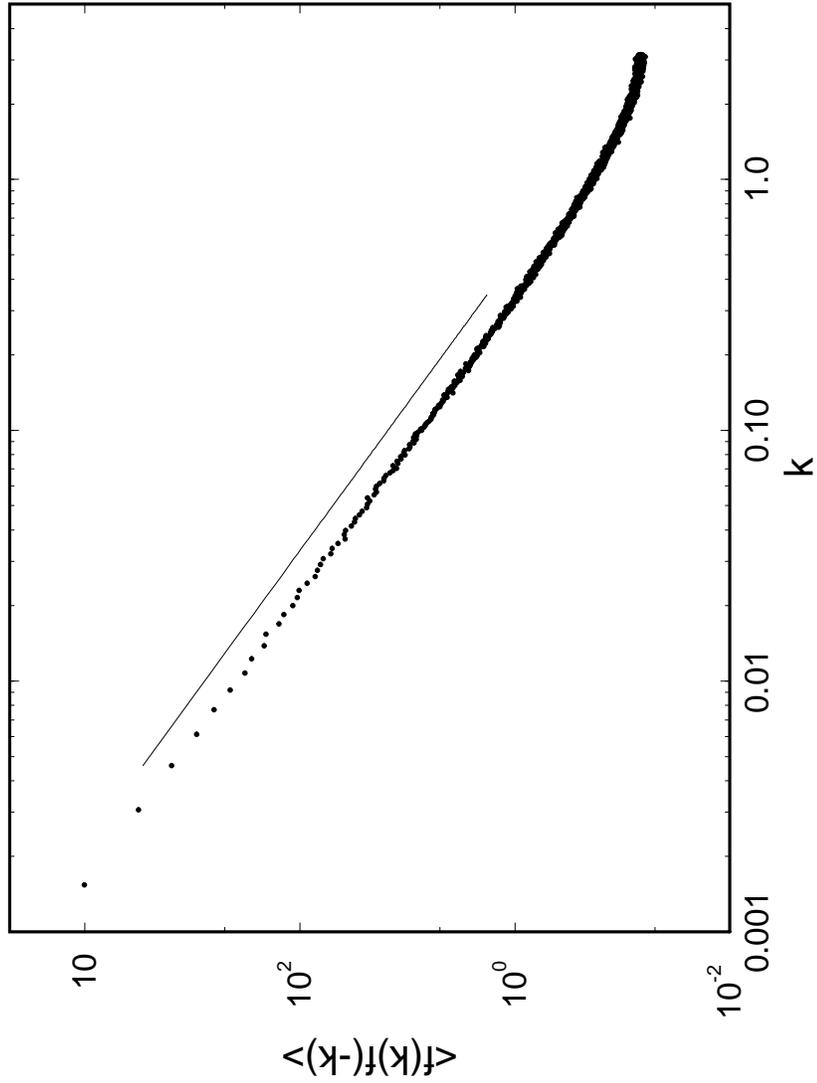}
\medskip
  \caption
{Log-log plot of the power spectrum for system size 4096, averaged over 
1000
samples and
a linear fit over $0.004\leq k \leq 0.35$ which leads
 to $2\zeta+1=1.68\pm 0.04$}   
  \label{zetaslope}
  \end{center}\end{figure}

\begin{figure}\begin{center}
\epsfxsize=5in
\epsffile{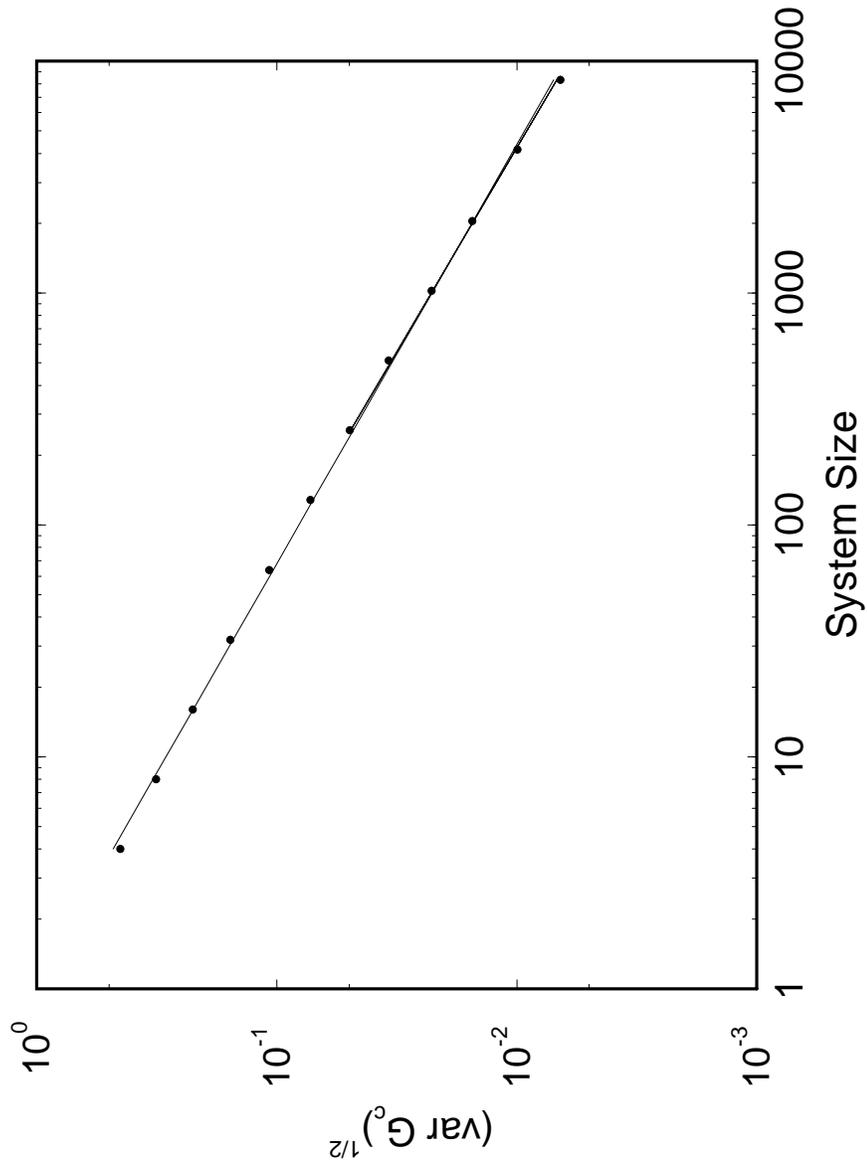}
\medskip
   \caption{Log-log plot of the variance of the quasistatic threshold
load as a function of the system size in the
 range from 4 to 8096. From a linear fit over the full range, 
 $\nu=1.80\pm0.05$ while a fit for sizes from 256 to 8096 yields 
$\nu=1.72\pm0.12$}   
  \label{nulinearfit}
  \end{center}\end{figure}

\begin{figure}\begin{center}
\epsfxsize=5in
\epsffile{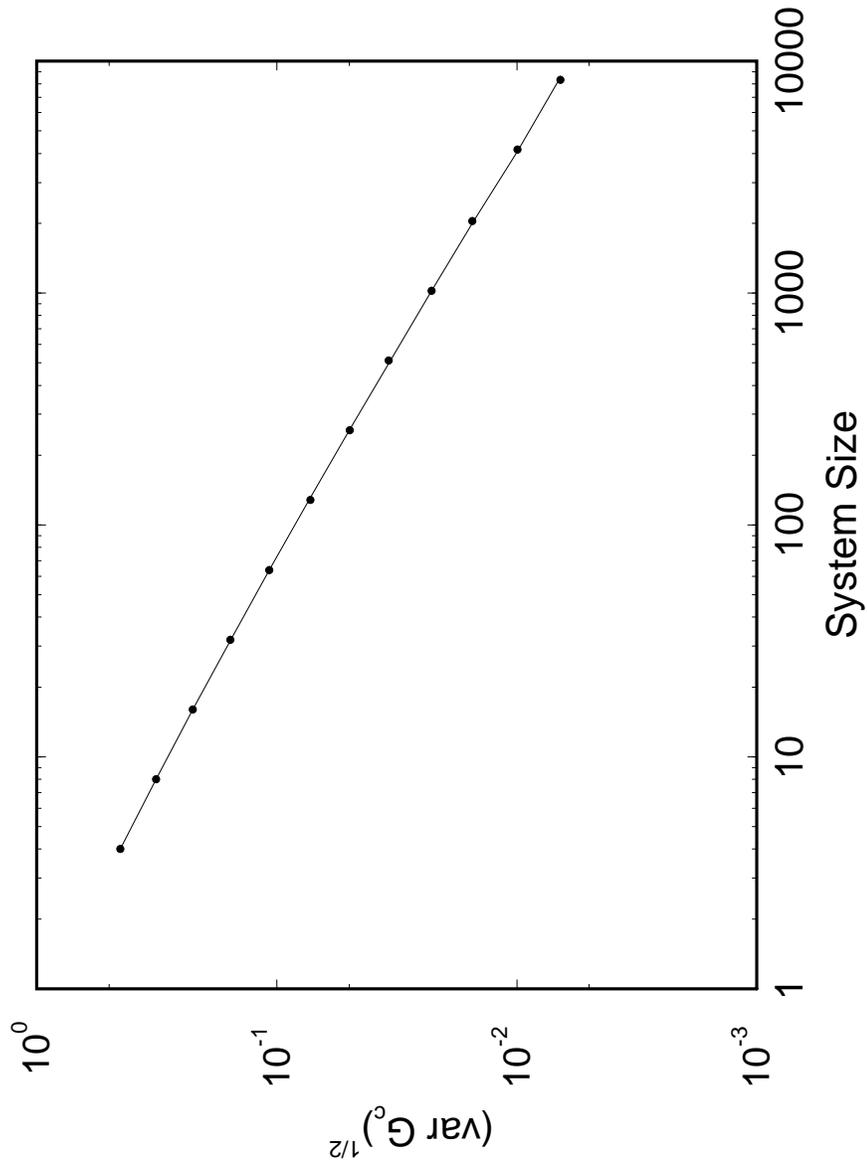}
\medskip
   \caption{Best fit of the variance of the quasistatic threshold
load as a function of the system size to the form $\frac{
CL^{-1/\nu }}{1+A_{\nu}L^{-1/3}}$. This yields
$\nu=1.52\pm0.02$ and $A_{\nu}=0.87\pm 0.03$}   
  \label{correctoscalingfit}
  \end{center}\end{figure}

\begin{figure}\begin{center}
\epsfxsize=5in
\epsffile{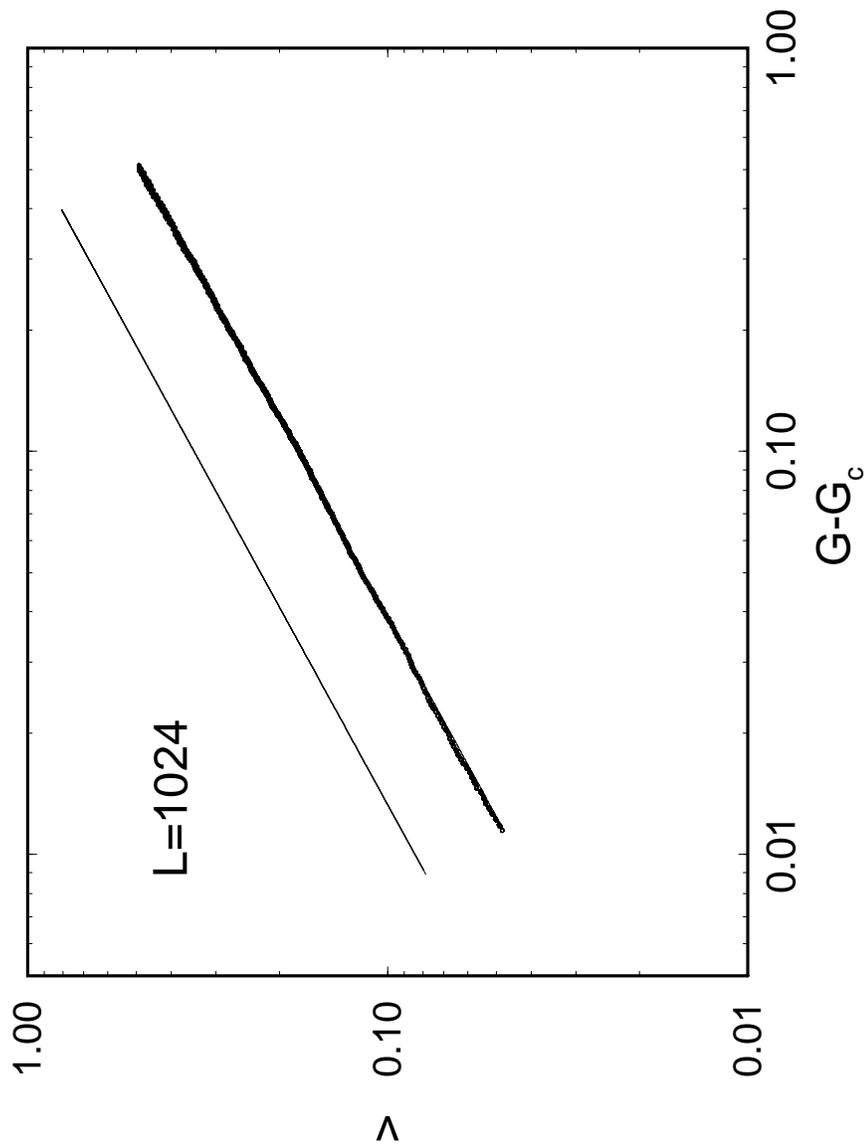}
\medskip
   \caption
{Log-log plot of the mean velocity in the quasistatic model
 versus the the excess of the load
above the critical value. The slope yields $\beta
=0.68\pm 0.06$.}   
  \label{beta}
 \end{center}\end{figure}

\begin{figure}\begin{center}
\epsfxsize=5in
\epsffile{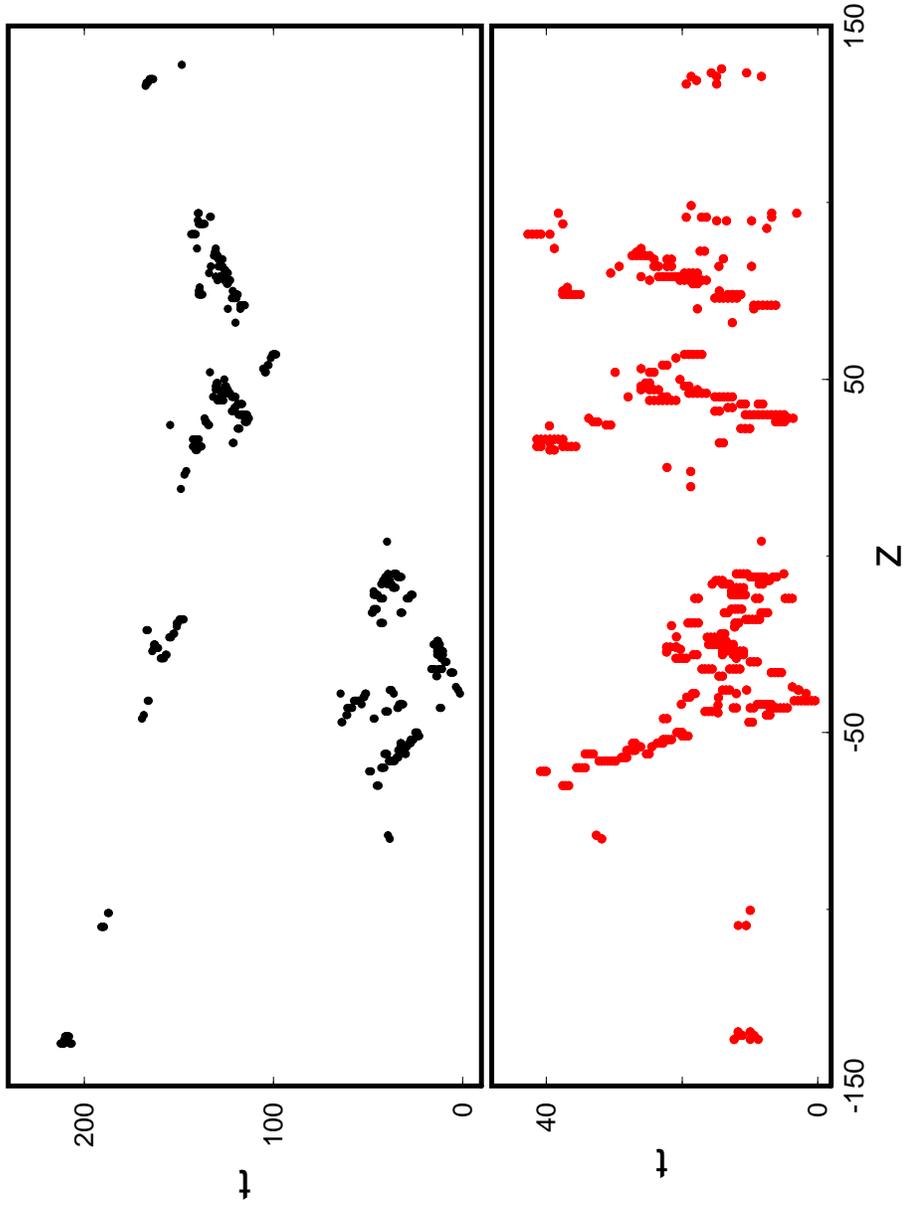}
\medskip
  \caption
{Space-time plot of an 
avalanche for (a) the time-delayed
monotonic model (top) and (b) for the quasistatic model(bottom)  starting 
from the same initial configuration of the crack front. The sets of jumps in 
the two
figures are identical, but they occur at different times}   
  \label{avmodqcomp}
  \end{center}\end{figure}

\begin{figure}\begin{center}
\epsfxsize=5in
\epsffile{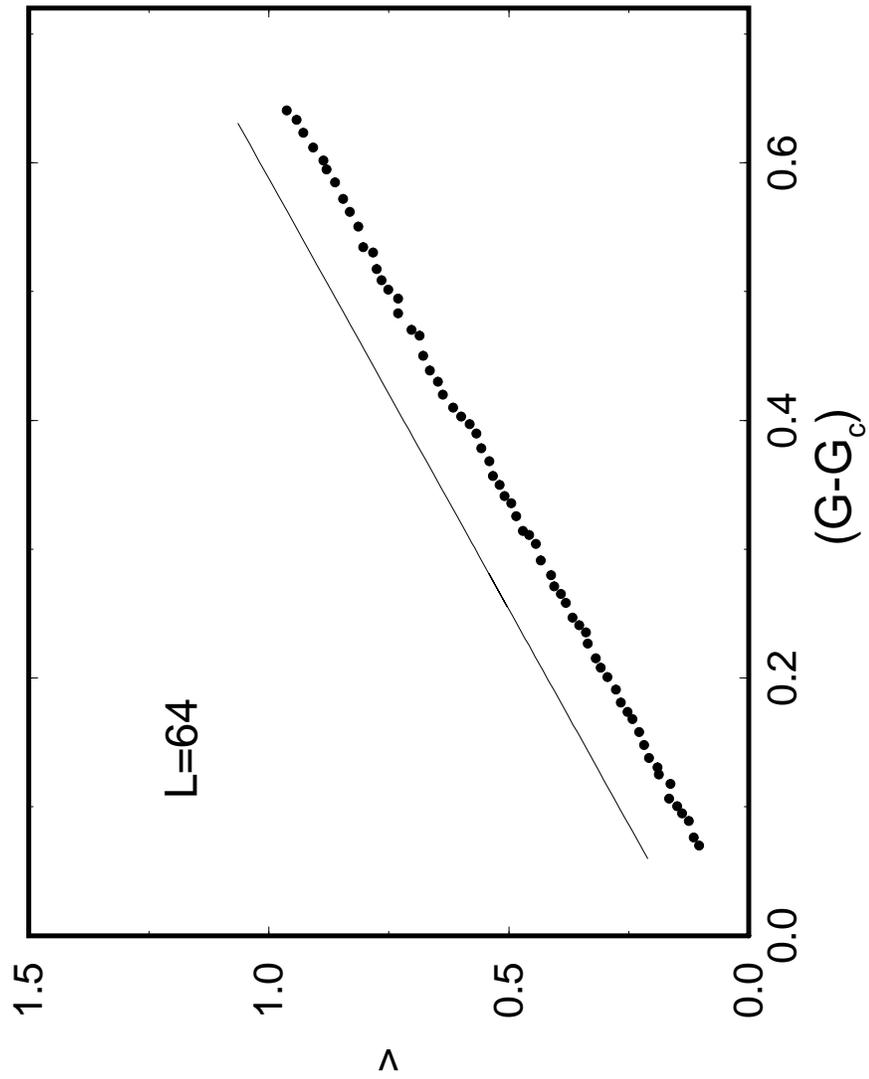}
\medskip
   \caption
{Velocity versus load for the time delayed
monotonic kernel in a system of size 64, and a linear fit to the data.}   
  \label{delayedmodqbeta}
  \end{center}\end{figure}

\begin{figure}\begin{center}
\epsfxsize=5in
\epsffile{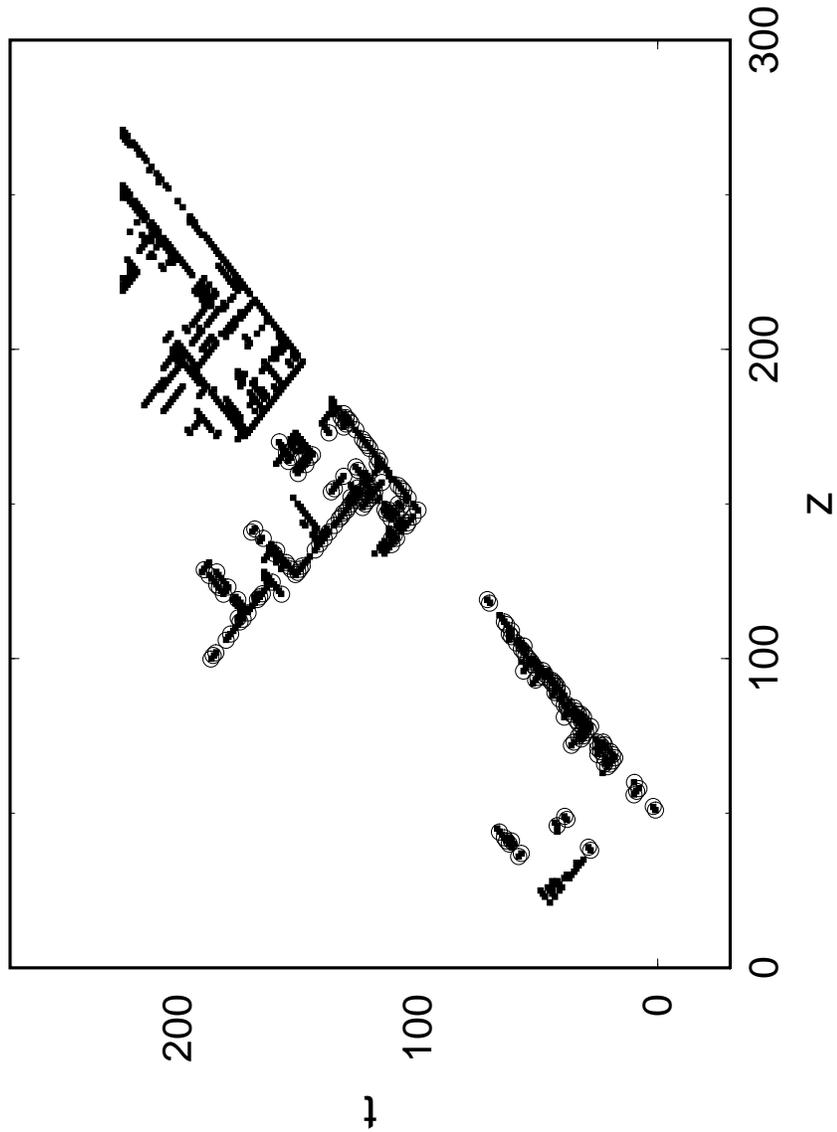}
\medskip
  \caption{Space-time plot of an avalanche for the time-delayed 
monotonic model (open circles) and
the stress pulse model with $\alpha=0.5,\gamma=1.5$ (with dots) from
identical initial conditions. It can be seen that a large number
of additional sites jump for the model with stress overshoots.}   
  \label{avpicovershoot}
  \end{center}\end{figure}

\begin{figure}\begin{center}
\epsfxsize=5in
\epsffile{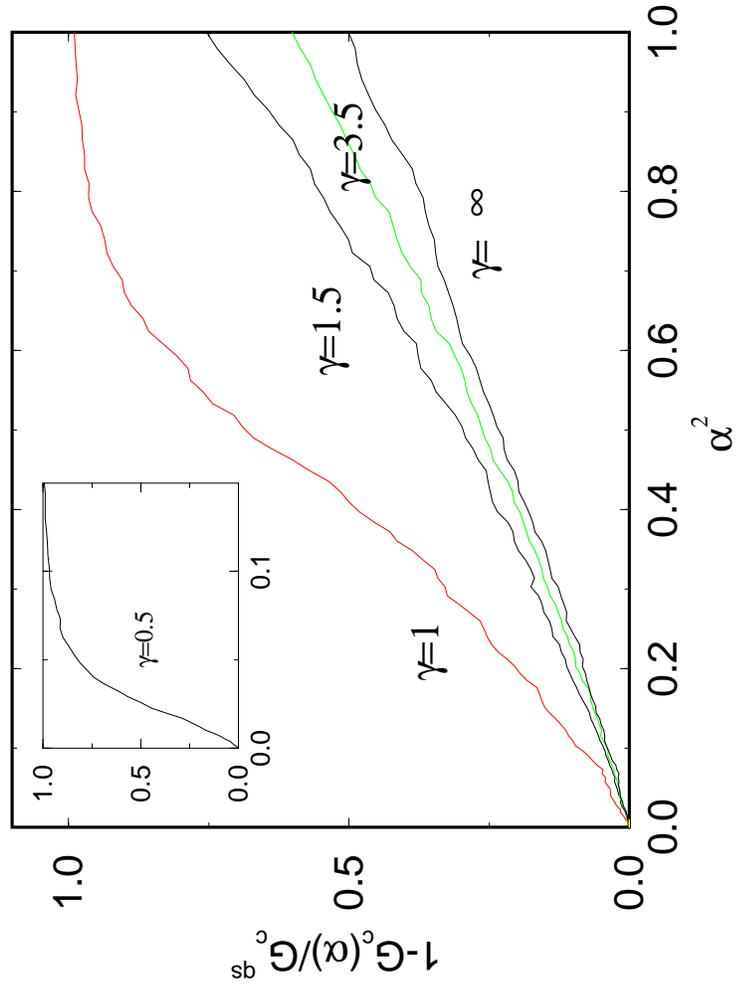}
\medskip
   \caption
{The fractional decrease in the threshold load, $1-G_{c}(\alpha ,\gamma )
/G_{c}^{qs}$ for a system of size
64, is plotted as a function of $\alpha^{2} $ for $\gamma =0.5,1,1.5.3.5$ 
and $
\infty $.}   
  \label{thresh}
 \end{center}\end{figure}

\begin{figure}\begin{center}
\epsfxsize=5in
\epsffile{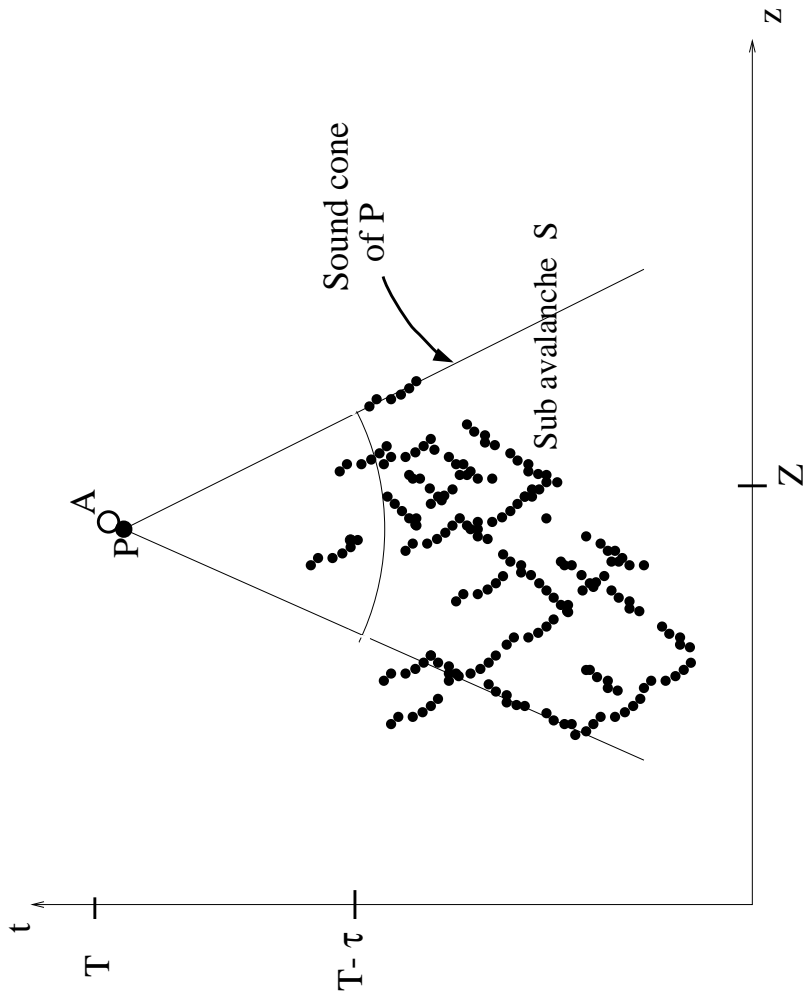}
\medskip
  \caption
{ Schematic of situation used in estimating the effects of a single small 
stress pulse. 
The space-time point $A$ at which we estimate the probability of an
additional jump due to a stress pulse from its neighbor jumping at a 
space-time point $P$, and $P$'s ``sound cone''
showing $P$'s isolation from the subavalanche, $S$, by a scale $\tau$, 
are all
 shown.}   
  \label{alphaargument}
  \end{center}\end{figure}

\begin{figure}\begin{center}
\epsfxsize=5in
\epsffile{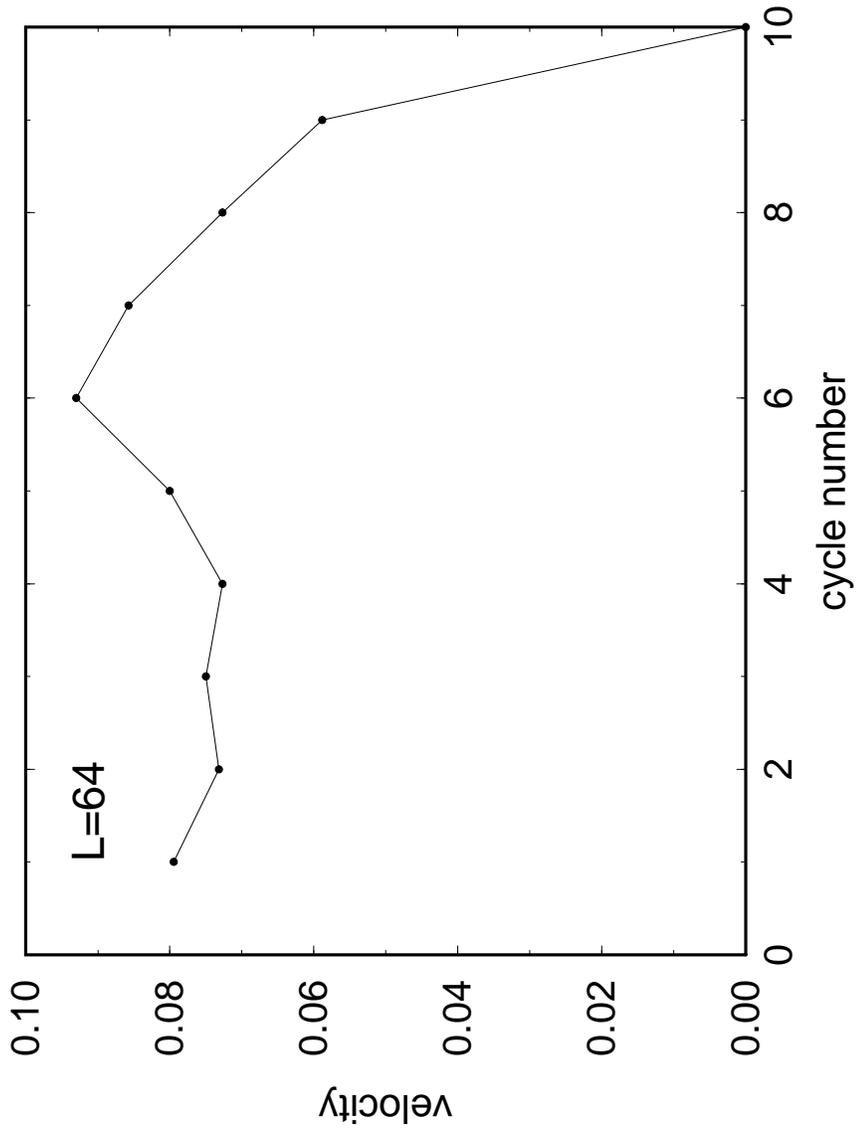}
\medskip
\caption
{ The velocity of the crack front for the stress pulse model with 
$\alpha=0.5,\gamma=1.5$
is shown as a function of cycle number for a system of size 64 and extent
16 with periodic
 boundary conditions in the direction of motion. The crack
front goes through the same random sample in each cycle and the
velocity averaged over each 
 cycle is measured while the external load is kept fixed.}   
  \label{abruptstop}
  \end{center}\end{figure}

\begin{figure}\begin{center}
\epsfxsize=5in
\epsffile{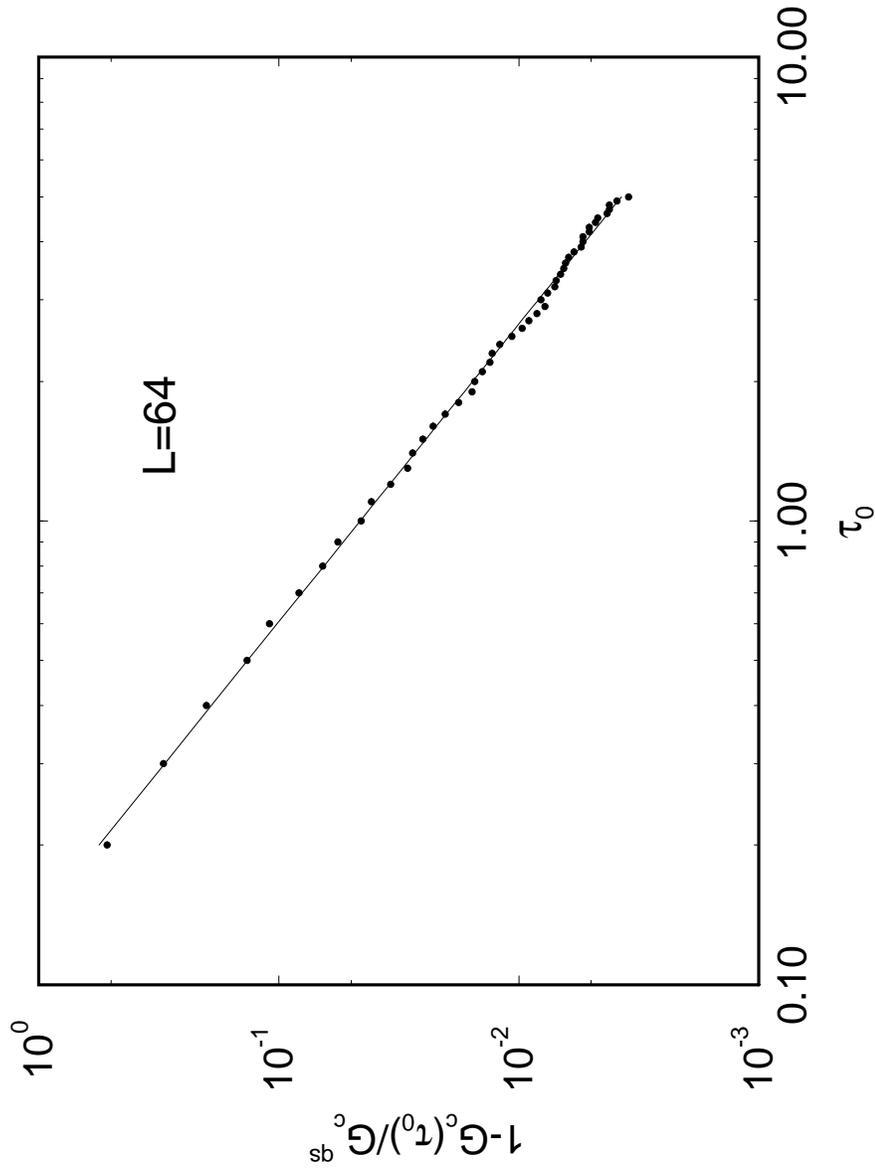}
\medskip
   \caption
{Log-log plot of the fractional decrease of the threshold force 
with scalar elastic stress transfer, $({
G_{c}^{qs}-G_{c}(\tau_{0} )})/G_{c}^{qs}$, versus $\tau_{0} $ is
shown for systems of size 128. 
The slope of the linear fit is $-1.56$}   
  \label{thforcewitho}
  \end{center}\end{figure}

\begin{figure}\begin{center}
\epsfxsize=5in
\epsffile{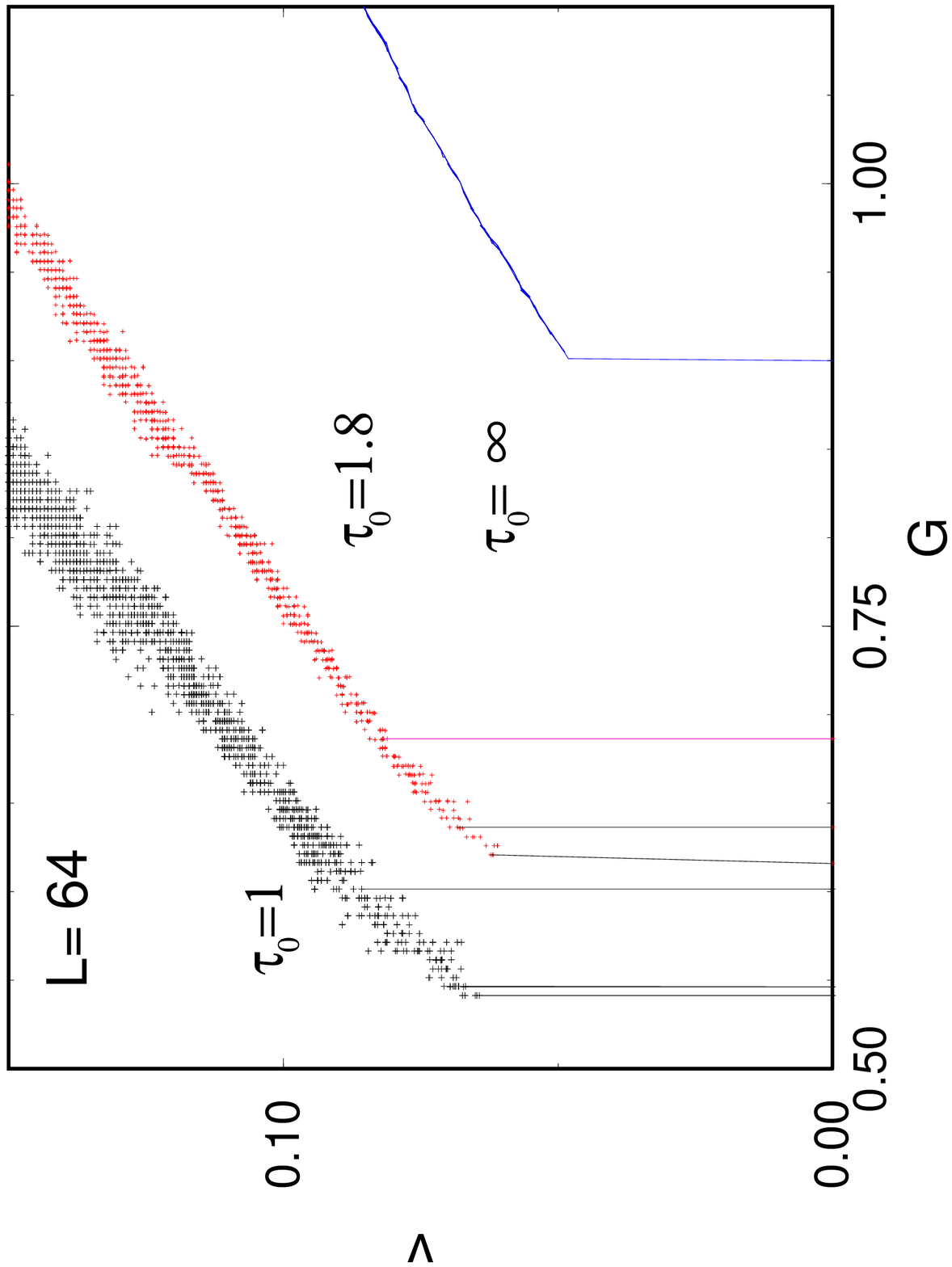}
\medskip
   \caption
{Mean velocity as a funtion of load for the scalar elastic
 model, in a single system of
size 64, for various values of $\tau_{0} $ ranging 
from 1 to $\infty $; $\tau_{0}
=\infty $ corresponds to the monotonic model with time delays 
where we expect a continuous
transition from the pinned to the moving phase. As can be seen, the finite
systems exhibit hysteretic behaviour.}   
  \label{hysteresis}
 \end{center}\end{figure}


\begin{references}
\bibitem{footnote2}  Note,
however, that in an ideal situation, if $G^{\infty }<\Gamma $ the crack will
close. In this case $v=A^{-1}(G^{\infty }/\Gamma )$ is negative and the
difference $v(1-A(v))G^{\infty }$, between the elastic energy absorbed by
the material and the energy released by healing goes into kinetic energy.
\end{references}
\end{document}